\documentclass[11pt]{article}

\usepackage{axodraw,graphicx,amsfonts,multicol,color}
\definecolor{rossos}{cmyk}{0,1,1,0.7}
\definecolor{blus}{cmyk}{1,1,0,0.7}

\def\bea{\begin{eqnarray}}
\def\eea{\end{eqnarray}}
\def\be{\begin{equation}}
\def\ee{\end{equation}}
\def\ba{\begin{array}}
\def\ea{\end{array}}
\def\nn{\nonumber}

\def\a{&\hspace{-6pt}}
\def\c{\hspace{-7pt}}
\def\gev{{\rm GeV}}
\def \lsim{\mathrel{\vcenter
     {\hbox{$<$}\nointerlineskip\hbox{$\sim$}}}}
\def \gsim{\mathrel{\vcenter
     {\hbox{$>$}\nointerlineskip\hbox{$\sim$}}}}

\def\circa#1{\,\raise.3ex\hbox{$#1$\kern-.75em\lower1ex\hbox{$\sim$}}\,}

\font\tenrsfs=rsfs10
\font\sevenrsfs=rsfs7
\font\fiversfs=rsfs5
\newfam\rsfsfam
\textfont\rsfsfam=\tenrsfs
\scriptfont\rsfsfam=\sevenrsfs
\scriptscriptfont\rsfsfam=\fiversfs
\def\mathscr#1{{\fam\rsfsfam\relax#1}}
\def\Lag{{\cal L}}

\begin{document}

\thispagestyle{empty}

\begin{center}
hep-th/0305184\hfill
\hfill CERN-TH/2003-079\hfill
\hfill IFUP-TH/2003-3\\

\begin{center}

\vspace{1.cm}

{\huge\bf Brane to brane gravity mediation

\vspace{3mm}

of supersymmetry breaking}

\end{center}

\vspace{1.cm}

{\large \bf Riccardo Rattazzi$^{a}$\footnote{On leave from INFN, Pisa, Italy.}, 
Claudio A. Scrucca$^{a}$, Alessandro Strumia$^{b}$}\\

\vspace{7mm}

${}^a$
{\em Theoretical Physics Division, CERN, CH-1211 Geneva 23, Switzerland}
\vspace{.2cm}

${}^b$
{\em Dipartimento di Fisica dell'Universit\`a di Pisa and INFN, Italy}
\vspace{.3cm}

\end{center}

\vspace{0.8cm}
\centerline{\bf Abstract}
\vspace{2 mm}
\begin{quote}
We extend the results of Mirabelli and Peskin to supergravity.
We study the compactification on $S^1/\mathbb{Z}_2$ of Zucker's
off-shell formulation of 5D supergravity and its coupling to matter at the fixed
points.
We clarify some issues related to the off-shell description of supersymmetry
breaking \`a la Scherk--Schwarz (here employed only as a technical tool)
discussing how to deal with singular gravitino
wave functions.

We then consider `visible' and `hidden' chiral
superfields localized at the two different fixed points and communicating only
through 5D supergravity. We compute the one-loop corrections that mix the two
sectors and the radion superfield. Locality in 5D ensures the calculability of
these effects, which transmit supersymmetry breaking from the hidden to the visible sector.
In the minimal set-up visible-sector scalars get a universal squared mass $m_0^2<0$.
In general (e.g.\ in presence of a sizable gravitational kinetic term localized on the hidden brane)
the radion-mediated contribution to $m_0^2$ can be positive and dominant.
Although we did not build a complete satisfactory model,
brane-to-brane effects can cure the tachyonic sleptons predicted by anomaly mediation
by adding a positive $m_0^2$ which is
universal up to subleading flavor-breaking corrections.
\end{quote}

\newpage

\setcounter{equation}{0}\renewcommand{\theequation}{\thesection.\arabic{equation}}

\section{Introduction}

In spite of the competition from other ingenious proposals, low energy supersymmetry 
remains the simplest and most realistic possibility for new physics at the 
electroweak scale. Among the reasons for that are its spectacular agreement with the 
expectations of Grand Unified Theories and its almost effortless satisfaction of the 
constraints posed by electroweak precision data. Nonetheless, at the theoretical level, 
there are still several unsatisfactory aspects, all directly related to the problem of 
supersymmetry breaking. Maybe the acutest problem is that supersymmetry should help 
with the cosmological constant problem, but it does not. Supersymmetry controls quantum 
corrections to the vacuum energy. However supersymmetry must be broken at or above the 
electroweak scale and the generic value of the cosmological constant is then 
$\gsim (100 \gev)^4$, an excess of at least fifty orders of magnitude.
In phenomenological applications of supersymmetry, the cosmological constant is tuned
to be small (at least it can be done!), with the hope that some other mechanism will 
explain that tuning. Another problem concerns the flavor structure of the squark and 
slepton mass matrices. This structure should be very specific in order to satisfy the 
experimental constraints on Flavor Changing Neutral Currents (FCNC). This requires 
theoretical control on the mechanism that generates the soft terms. Finally, the Higgs 
sector and electroweak symmetry breaking are crucially controlled by the $\mu$-parameter, which 
does not itself break supersymmetry. The special status of $\mu$ compared to the other
mass terms, which do break supersymmetry, is often a serious obstacle to the construction 
of simple and realistic theories for the soft terms. Indeed, after the completion of the 
LEP/SLC program, without the discovery of any superparticle, there is yet another source 
of embarrassment for supersymmetry: why is supersymmetry hiding in experiments at the weak 
scale if its role is to explain the weak scale itself? Quantitatively: with the present 
lower bounds on the sparticle masses the reproduction of the measured $Z$-mass 
requires a fine-tuning of at least $1/20$ among the parameters of all popular models. 
Basically, more than $95 \%$ of their parameter space is already ruled out. If we want 
to stick to supersymmetry, is there a message in the need for this tuning? Is it possible 
that this tuning is not accidental, and that the underlying model naturally selects somewhat 
heavier than expected sparticle masses?

All in all the above problems are probably telling us that we have not yet a fully 
realistic model for the soft terms. The hope and the assumption in the quest for such 
a model is usually that the first problem, the cosmological constant problem, due to his 
hugely different nature, will find a separate solution, not affecting physics at the 
weak scale. In this paper we will follow this standard path and concentrate of the 
flavor problem.

\begin{figure}[t]
\vskip 20pt
\begin{center} 
\includegraphics[width=8.5cm,height=5.2cm]{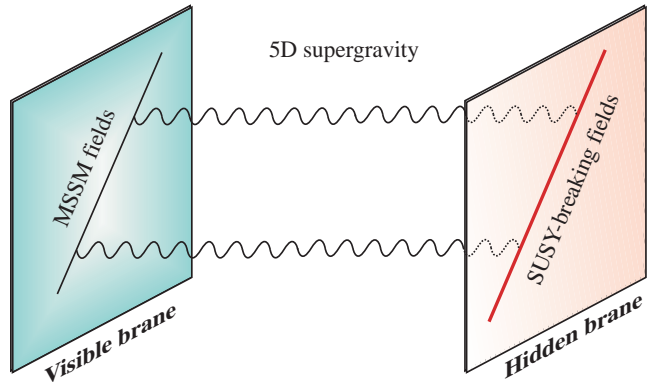}\\
\caption{\em One-loop supergravity diagrams inducing an effective interaction between 
visible and hidden sector.}
\label{b2bmed}
\end{center}
\end{figure}

In the Standard Model (SM) all flavor violation arises in the fermion mass matrices themselves. 
FCNC are then naturally suppressed, in agreement with experimental data, by powers of the fermion 
masses and mixing angles. This is the Glashow-Ilipoulos-Maiani (GIM) mechanism. 
In the Minimal Supersymmetric Standard Model 
(MSSM) a generic sfermion mass matrix represents a new source of flavor mixing, not aligned with 
the fermion mass matrices. The GIM mechanism generically does not work in the MSSM, and FCNC bounds 
are not satisfied. A model for the soft terms enforcing the GIM mechanism would tackle this 
difficulty. Gauge mediated models~\cite{Dine} (see~\cite{Gaugemed} for a review) are 
such an example. In that case soft terms are mediated by gauge interactions at a scale 
$M$ much below the flavor scale $\Lambda_F$. The resulting soft terms are flavor symmetric 
up to small effects due to the SM Yukawa matrices themselves. Extra dangerous flavor 
violating effects are further suppressed by powers of $M/\Lambda_F$. The resulting 
FCNC are then analogous to those of the SM. Gauge mediated models are very attractive 
in this respect, but they require extra inelegant complication to solve the $\mu$-problem. 
The so-called gravity mediated models~\cite{Gravmed}, on the other hand, fare better 
on the $\mu$-problem (thanks to the possibility of the Giudice--Masiero mechanism 
\cite{GiudiceMasiero}) but are in trouble with flavor. 
At first this seems surprising since gravity is as flavor universal as the SM gauge interactions. 
However the point is that gravity is universal, or more precisely it respects GIM, only in the 
IR. On the other hand, gravity mediated models effectively represent the generation of soft 
masses by UV phenomena in the fundamental theory of quantum gravity. Now, this unknown fundamental 
theory has to explain why the top quark is so much heavier than the up quark and everything 
else: it should also be the theory of flavor. Then it is not obvious why it should generate
soft terms respecting the GIM mechanism. The presence of extra-dimensions can however change 
this state of affairs. The key is a new scale associated to the radius of compactification $R$. 
The prototypical example 
is provided by the ``sequestered sector'' scenario suggested by Randall and Sundrum~\cite{Randall},
and inspired by  string~\cite{Dixon} and $M$-theory orbifolds~\cite{HW}
(although it seems difficult to realize this scenario in string models~\cite{Anis}).
The model involves one extra dimension compactified on the orbifold $S_1/\mathbb{Z}_2$. The MSSM 
lives at one boundary, say $x^5=0$, while the supersymmetry breaking sector 
lives at the other boundary, a distance $\pi R$ away. It is assumed that $R$ is parametrically 
bigger than the 5D Planck length $1/M_5$. Locality in 5D insures the absence of direct tree 
level couplings between the two sectors \cite{ChackoLuty}.
The direct uncalculable couplings were the origin 
of flavor violation in ordinary 4D models. At the quantum level the two sectors couple through 
virtual graviton exchange, see Fig.~\ref{b2bmed}. These loops are saturated at virtuality 
$\sim 1/\pi R$: as long as $R\gg 1/M_5$, they are dominated by the IR flavor universal regime 
of gravity.\footnote{For example, extra particles with mass $M\sim M_5$ propagating in the extra dimension
might be present in a fundamental theory of gravity, giving extra
contributions suppressed by $\exp(- MR)$.}
Indicating by $F_{\Phi_\pi}\sim M_{\rm susy}^2$ the Vacuum Expectation Value 
(VEV) of the auxiliary fields in the hidden sector, at 1-loop the universal scalar mass $m_0$ 
is of order. 
\be
m_0^2\sim \frac{1}{16\pi^2}\frac{|F_{\Phi_\pi}^2|}{M_5^6 (\pi R)^4} \;.
\label{b2b}
\ee
This effect was never computed so far. The reason is that for $R M_5\to \infty $ the leading 
contribution to soft terms comes from another quantum effect, where gravity enters only at the 
classical level:  the so called anomaly mediated supersymmetry breaking (AMSB) 
\cite{Randall,Giudice}. The auxiliary field $F_{S_0}$ acting as a source in AMSB is the one 
in the gravitational supermultiplet 4D Poincar\'e supergravity.
This field couples to the MSSM only via the superconformal anomaly. Being an anomaly, 
this effect is completely saturated in the IR. Again, only the universal aspects of gravity 
play a r\^ole, and the anomaly mediated sfermion masses beautifully enforce the GIM mechanism. 

\smallskip

Unfortunately, the sleptons turn out to be tachyonic, as a sharp consequence of ${\rm SU}(2)_L$ 
not being asymptotically free in the MSSM. Moreover the $\mu$-problem affects AMSB very much 
as GMSB. Various proposals have been made to fix these problems.
Indeed if one assumes that some unspecified flavor universal contribution $m_0^2>0$ lifts 
the sleptons, then the low-energy phenomenology is quite peculiar~\cite{AM+m0}. 
The main purpose of this paper is to study whether and how the brane-to-brane mediated 
term in eq.~(\ref{b2b}) can realize this situation.

The anomaly mediated gaugino masses $m_{1/2}$ and scalar masses $m_s$ depend on the 
auxiliary scalar $F_{S_0}$ of supergravity and scale roughly like
\be
\label{AM}
m_{1/2} \sim m_s\sim \frac{g^2}{16\pi^2}\, |F_{S_0}| \;.
%\;,\quad\quad\quad m_s^2 \sim \frac{g^4}{(16\pi^2)^2}\, |F_{S_0}|^2 \;.
\ee
In the minimal situation, $F_{S_0} \sim F_{\Phi_\pi}/M_4$ where the 4D reduced Planck mass 
$M_4$ is defined as $M_4^2 = M_5^3 \pi R$. Although AMSB scalar masses squared arise at two-loop, 
they dominate eq.~(\ref{b2b}) for $(M_5 \pi R)^3\gsim 16\pi^2=(M_5\pi R_{\rm cr})^3$ (na\"{\i}ve  
dimensional analysis~\cite{nda} estimates that quantum gravity effects become important around 
or below the energy $\Lambda_5 \sim 4\pi M_5$). If the radius were stabilized at the critical 
value $R_{\rm cr}$, and if the brane-to-brane contribution were positive, the tachyon problem 
could be overcome while preserving a certain control on flavor universality. Notice indeed that 
$R_{\rm cr}$ is still parametrically larger that the Planck length. Notice also that gaugino 
masses are not affected by the brane-to-brane loops. Therefore, if $R<R_{\rm cr}$ gauginos are 
parametrically lighter then sfermions, which requires extra fine tuning in electroweak breaking. 
In ref.~\cite{Luty} a simple mechanism of radius stabilization which can plausibly give 
$R\sim R_{\rm cr}$ was pointed out. 

As we said, the purpose of the present paper is to calculate the brane-to-brane term $m_0^2$.
In fact we will do more and calculate the full 1-loop correction to the K\"ahler potential, 
or better its IR saturated part. Along the way, we will also study in some detail the interactions 
of boundary fields with bulk supergravity. The paper is organized as follows. In the next section 
we outline the strategy that we will use to perform our computation. In section 3 we discuss 
the Lagrangian for off-shell 5D supergravity and its coupling to the boundary. In section 4 
we show in a sample computation that supergravity cancellation are correctly reproduced. 
Section 5 is a detailed discussion of Scherk--Schwarz supersymmetry breaking, that we will
need only as a tool to compute the 1-loop correction to the K\"ahler potential. 
In our phenomenological applications supersymmetry is not broken just by the Scherk--Schwarz 
mechanism. In section 6 we present our computation. In section 7 we discuss our results and their 
consequences. Finally, section 8 is devoted to conclusions.

\setcounter{equation}{0}
\section{Outline}
\label{outline}

In this section, we will describe the general context in which we will work and outline
the main steps of the computation that we will perform.

\subsection{The model}

We consider a 5D supergravity model compactified on $S_1/\mathbb{Z}_2$, following closely the study by 
Luty and Sundrum~\cite{Luty}. We parametrize $S_1/\mathbb{Z}_2$ by $x^5\simeq x^5 +2\pi$ and $x^5\simeq -x^5$.
We assume that all the fields of the MSSM live at $x^5=0$, while at $x^5=\pi$ there is a field 
theory breaking supersymmetry in the flat limit, the hidden sector. For the purpose 
of our calculation it is enough to consider a toy MSSM consisting of just one chiral superfield 
$\Phi_0$ (containing a sfermion $\phi_0$, a Weyl fermion $\chi_0$ and the auxiliary field $F_{\Phi_0}$). 
The result for the MSSM will just be a straightforward generalization. Similarly we assume that 
the hidden sector is effectively described by an O'Raifertaigh model involving just one chiral 
superfield $\Phi_\pi$. We will assume that all interactions in the hidden sector are 
characterized by just one scale $\Lambda$, corresponding to its interpretation as the low energy 
description of a dynamical supersymmetry breaking model. Since the radius $R$ is also a massless
field in the lowest order description of the scenario, we will have to include it in the effective
4D description and  to determine the vacuum dynamics. At low energy the effective tree level 
(classical) K\"ahler function $K_{\rm cl} = - \frac 32\, {\rm ln} [-\frac 23 \Omega_{\rm cl}]$ 
is then specified by
\be
\Omega_{\rm cl} = - \frac {3}{2} (T + T^\dagger) M_5^3 
+ \Omega_0(\Phi_0,\Phi_0^\dagger) + \Omega_\pi(\Phi_\pi,\Phi_\pi^\dagger)
\label{Omega0}
\ee
where $T$ is the radion superfield, and $\Omega_0$ and $\Omega_\pi$ are 
the contributions to the gravitational kinetic function coming respectively from the $0$ and $\pi$ 
fixed points. The gravitational action is proportional to the D-term $[\Omega_{\rm cl}S_0S_0^\dagger]_D$, 
where $S_0$ is the chiral compensator\footnote{We are using here the superconformal formulation of the 
effective 4D theory \cite{supconf}.}. By the above additive  form of $\Omega_{\rm cl}$ we have that 
the VEV of the auxiliary fields $F_T$ and $F_{\Phi_\pi}$ do not generate soft terms in the visible 
sector. (Notice however that in the Einstein frame the two sector have mixed kinetic terms.) 
At this stage the visible sector soft terms are generated through anomaly mediation and are 
proportional to $F_{S_0}\sim m_{3/2}$. In this paper we will calculate the 1-loop correction 
$\Delta \Omega$ to $\Omega_{\rm cl}$, which introduces direct coupling between visible, hidden and 
radion sectors. The soft terms generated by $\Delta \Omega$ depend on $F_{\Phi_\pi}$, $F_T$ and 
$T$. The relations among these parameters are strongly dependent on the mechanism that stabilizes 
$T$.

\subsection{The computation} 

We now illustrate our strategy to compute the 1-loop correction $\Delta\Omega$. The first 
remark is that, like  $\Omega_{\rm cl}$,  $\Delta \Omega$ must depend on $T$ and $T^\dagger$ 
only through the combination $T+T^\dagger$ whose lowest component is the length $\pi R$ of the 
internal dimension. The reason is that the lowest component of $T-T^\dagger$ is the internal 
component of the graviphoton $A_5$, which couples only derivatively in the tree level Lagrangian. 
A dependence of $\Delta \Omega$ on $T - T^\dagger$ would lead to non derivative terms in $A_5$, 
which cannot happen in perturbation theory\footnote{At the non perturbative level, these terms 
can be generated, via for instance instanton effects, like in eq.~(\ref{gaugecond}).}. 
So $\Delta \Omega = \Delta\Omega(T+T^\dagger,\Phi_{0,\pi},\Phi_{0,\pi}^\dagger)$.

We calculate $\Delta \Omega$ by a little trick: we reconstruct it by computing the 1-loop effective 
scalar potential $\Delta V$ induced in a background with $F_T \neq 0$ and with all other auxiliary 
fields vanishing. This scenario is consistently realized in our model if a constant boundary 
superpotential $P=c$ is chosen. At tree level this is the simple suspersymmetry breaking no-scale 
model~\cite{noscale1} (see~\cite{noscale2} for a review): $F_T\sim c$ and $F_{S_0}=0$, where the 
second condition ensures exact cancellation of the effective cosmological constant. In our 5D model 
this way of breaking supersymmetry is completely equivalent to the Scherk--Schwarz mechanism 
\cite{Scherk}. In section 5, to clarify our procedure, we will have to take a detour into 
explaining in detail the relation to the Scherk--Schwarz mechanism. Now, at zero momentum we have
\be
\Delta V =-\left[\Delta \Omega S_0S_0^\dagger\right]_D
=-|F_T|^2 \partial_T \partial_{T^\dagger} \Delta \Omega(T+T^*,\phi_{0,\pi},\phi_{0,\pi}^*) \;,
\label{relation}
\ee
where $\Delta V$ is the quantity we calculate, with $F_T$ as an input.
Eq.~(\ref{relation}) is a simple differential equation whose solution 
gives $\Delta \Omega$ up to two integration ``constants'' $H_0$ and $H_1$:
\be
\Delta \Omega=\Delta \hat \Omega(T+T^*,\phi_{0,\pi},\phi_{0,\pi}^*)
+ H_0(\phi_{0,\pi},\phi_{0,\pi}^*)
+(T+T^*)H_1(\phi_{0,\pi},\phi_{0,\pi}^*) \;.
\ee
The quantity $\Delta\hat \Omega$ is entirely determined and explicitly anticipated below.
On the other hand, the form of the unknown $H_{0,1}$ is strongly constrained by 5D 
locality and the limit $R\to \infty$. Since $\Phi_0$ and $\Phi_\pi$ are located at the 
two different boundaries and cannot talk to each other in the limit $R\to \infty$, $H_0$ 
must have the form:
\be
H_0=\Delta\Omega_0(\Phi_0,\Phi_0^\dagger)+\Delta\Omega_\pi(\Phi_\pi,\Phi_\pi^\dagger) \;.
\ee
Then it is clear that $H_0$ is just associated to the local, UV divergent, renormalization 
of each boundary kinetic function, and does not contribute to brane to brane mediation of 
supersymmetry breaking. $H_1$ is  an ``extensive'' contribution, growing with the volume 
and must be associated to renormalization of local bulk operators. Therefore $H_1$ cannot 
depend on the boundary fields: it is a constant associated to the uncalculable renormalization 
of the 5D Planck mass. So the only relevant quantity is the calculable one, $\Delta \hat \Omega$. 
\footnote{This discussion, although correct, needs an extra remark to be made fully rigorous. 
This is because the one dimensional Green function grows linearly with the separation and 
contributions that are linear in $T$ and mix the fields at the two boundaries are in principle 
possible. Indeed such an effect arises at tree level from the exchange of one graviton. 
However it corresponds to a 4-derivative interaction in the effective theory \cite{ChackoLuty}, 
and so it does not concern us. Now the basic point is that at the quantum level we are considering 
1-PI diagrams, where at least two gravitons are exchanged between each boundary: these diagrams 
have at least one further suppression $1/(M_5T)^3$, so that their contribution vanishes at least 
as $1/T^2$ for $T\to \infty$. In fact for two derivative operators (K\"ahler) there is an extra 
$1/T^2$ suppression by simple dimensional analysis, see eq.~(\ref{DeltaOmegalow}).}

\smallskip

The computation of $\Delta V$ requires in principle the knowledge of all the interactions 
between the boundary matter fields and the bulk supergravity fields. These can be obtained 
from the ordinary 4D supergravity tensor calculus, once the boundary values of bulk fields 
have been appropriately combined into 4D supermultiplets. We will do this in some detail 
in section \ref{sugra}, by using the off-shell description of 5D supergravity developed in 
\cite{Zucker}, thereby extending the results of~\cite{Mirabelli} from global to local supersymmetry.
Our results do not fully agree with previous attempts (for instance \cite{Gherghetta}), 
and we therefore verify them in section~\ref{test} by checking that the basic cancellations 
demanded by supersymmetry are reproduced. Computing $\Delta V$ turns out to be an easy task. 
Since it vanishes in the 
supersymmetric limit $F_T=0$, and since only the mass spectrum of gravitinos is affected by 
a $F_T\neq 0$, $\Delta V$ is simply given by the gravitino loop contribution, minus its value 
for $F_T=0$ (the same remark was used in \cite{Gherghetta}).
Furthermore, as a consequence of being in 5D and working at zero momentum, the 
whole contribution comes from diagrams involving only the scalar-scalar-gravitino-gravitino 
coupling. Such couplings are the same as those occurring in 4D supergravity.

\medskip

For our phenomenological applications it is enough to consider the following form of the 
boundary kinetic functions
\bea
\Omega_0 \a=\a - 3 L_0 M_5^3 + \Phi_0 \Phi_0^\dagger \;,
\label{lowest0} \\
\Omega_\pi \a=\a - 3 L_\pi M_5^3 + \Phi_\pi \Phi_\pi^\dagger \;.
\label{lowestpi}
\eea
The constants $L_{0,\pi}$ represent localized kinetic terms for the bulk supergravity fields, 
like those considered for pure gravity in ref.~\cite{dgp}. Negative values of $\Omega_{0,\pi}$ 
correspond to positive kinetic terms. For $\Omega_0$, the above form is motivated by the fact 
that for phenomenological applications we can work close to the origin in field space. We do not 
consider a linear term in $\Phi_0$ since there are no gauge singlets in the MSSM. In the hidden 
sector, we can always choose $\Phi_\pi$ such that the VEV of $\phi_\pi$ vanishes. Then a  simple 
analysis shows that terms of cubic and higher order do not contribute to soft terms in the 1-loop 
approximation. In general, there will however be a linear term in $\Phi_\pi$, which corresponds to 
$\Phi_\pi \to \Phi_\pi + {\rm const}$ in eq.~(\ref{lowestpi}).

\medskip

Let us conclude this section by anticipating our main result. We find that the calculable 
1-loop correction $\Delta \hat \Omega$ to the K\"ahler potential is given by 
\bea
\Delta \hat \Omega \a=\a -\frac {9}{\pi^2} M_5^2 \int_0^\infty \!\!\!\! dx\, x\, 
\hspace{1pt}{\rm ln} \Bigg[1 - \frac {1 + x\, \Omega_0 M_5^{-2}}{1 - x\, \Omega_0 M_5^{-2}}\,
\frac {1 + x\, \Omega_\pi M_5^{-2}}{1 - x\, \Omega_\pi M_5^{-2}}\,
e^{-6x(T + T^\dagger)M_5}\Bigg] \;.
\label{DeltaOmega}
\eea
We believe that this result is valid for general $\Omega_{0,\pi}$, and not just those in 
eqs.~(\ref{lowest0}) and (\ref{lowestpi}), but to prove this rigorously would require some 
more precise discussion into which we will not enter. In the standard situation $L_{0,\pi}=0$
(or negligibly small), expanding at the lowest order in $\Phi_0$ and $\Phi_\pi$ we find
\be
\Delta \hat \Omega = \frac{\zeta(3)}{4\pi^2(T+T^\dagger)^2} 
+ \frac{\zeta(3)}{6\pi^2} \frac {\Phi_0 \Phi_0^\dagger + \Phi_\pi \Phi_\pi^\dagger}
{(T + T^\dagger)^3 M_5^3} + \frac {\zeta(3)}{6\pi^2}\, \frac {\Phi_0 \Phi_0^\dagger 
\Phi_\pi \Phi_\pi^\dagger}{(T + T^\dagger)^4 M_5^6}
+ \cdots \;. 
\label{DeltaOmegalow}
\ee
The first term in (\ref{DeltaOmegalow}) is the well known Casimir energy correction.
The third  term gives {\em brane-to-brane mediation} of SUSY breaking.
The second term induces {\em radion-mediated\/} SUSY breaking, if the 
radion field $T$ also gets a non-zero $F$ term ($F_T$ has dimension zero).
It was previously computed in~\cite{Gherghetta}, and we agree with their 
result. The  order of magnitude of the coefficients agree with a na\"{\i}ve 
estimate performed in the effective 4D theory, where these terms are UV 
divergent, with a cut-off $\Lambda_{\rm UV}\sim 1/\pi R$. 

\setcounter{equation}{0}
\section{Full five-dimensional theory}
\label{sugra}

In this section we consider 5D supergravity compactified on $S^1/\mathbb{Z}_2$, 
with 4D chiral and vector multiplets localized at the two fixed points $x^5=0$ 
and $x^5=\pi $, which we will refer to as respectively the visible and the hidden branes. 
Our aim is to write the couplings between bulk and brane fields. This can be 
done by working with an off-shell formulation of supergravity as done in~\cite{Mirabelli} 
for the simpler case of rigid  supersymmetry. Our discussion is based on the work
of Zucker~\cite{Zucker}, in which  both the 5D off-shell Lagrangian
and the projected multiplets at the boundary were derived.

A few words on notation are in order. We set $M_5=1$. 
We use Latin capitals $A,B,\dots = \dot{1},\dots,\dot{5}$ 
for the flat 5D space time indices and Latin capitals from the middle alphabet 
$M,N,\dots=1,\dots,5$ for the curved 5D indices. Similarly we use 
$\alpha,\beta,\dots=\dot{1},\dots,\dot{4}$  for the flat 4D indices and 
$\mu,\nu,\dots=1,\dots, 4$ for the 4D curved ones. 
The 5D fermions are simplectic Majorana spinors, and carry ${\rm SU}(2)_R$ indices denoted with 
$i,j,\dots$; they satisfy the condition $\bar \Psi_i = \varepsilon_{ij}\Psi^{j T} C$, 
where $C$ is the charge conjugation matrix, and can thus be decomposed in terms of two 
Weyl spinors $\chi^i$ as follows: $\Psi^i = (\chi^i, \varepsilon^{ij} \bar \chi_j)^T$. 
As usual, the Weyl spinors $\chi^i$ can be equivalently described in terms of Majorana 
spinors $\psi^i = (\chi^i, \bar \chi^i)^{T}$. Occasionally we shall also use the 
${\rm SU}(2)_R$ doublet of Weyl spinors $\chi = (\chi^1,\chi^2)^T$.
Our conventions are such that $\gamma^{\dot 5} = {\rm diag}(-i,-i,i,i)$ and 
$\varepsilon^{12}=1$.

Consider first the bulk theory on $S^1$. The on-shell version contains the 
f\"unfbein $e_M^A$, the gravitino $\Psi_M^i$ and the graviphoton $A_M$, and 
has a global ${\rm {\rm SU}(2)}_R$ symmetry under which the gravitino is a doublet
\cite{Gunaydin}. Its minimal off-shell extension has been described in~\cite{Zucker}. 
It involves a minimal supergravity multiplet $(e^A_M,\Psi_M,A_M;\vec t, v_{AB}, 
\vec V_M, \lambda,C)$ containing the physical degrees of freedom and a set of 
auxiliary fields, where we indicate by an upper arrow the ${\rm SU}(2)_R$ triplets. 
In particular $\vec V_M$ gauges the ${\rm {\rm SU}(2)}_R$ symmetry. In addition, there 
is a compensator multiplet containing only auxiliary fields. The most convenient choice is a tensor 
multiplet $(\vec Y, B_{MNP}, \rho, N)$, which is related to a linear multiplet in 
which the constraint is solved by Poincar\'e duality with a vector component defined as 
\be
W^M = \frac 1{12}\,\epsilon^{MNPQR} \partial_N B_{PQR} 
+ \frac 14 \bar \Psi_P \vec \tau \gamma^{PMQ} \Psi_{Q} \vec Y
- \frac i2 \bar \rho \gamma^{MN} \Psi_N \;.
\label{WM}
\ee

The theory on $S^1/\mathbb{Z}_2$ is defined by assigning each field a $\mathbb{Z}_2$ parity such 
that the Lagrangian is an even density. The orbifold projection then globally breaks 
${\cal N}=2$ down to ${\cal N}=1$ and ${\rm SU}(2)_R$ down to $\hbox{U}(1)$. There is a two parameter 
family of possible choices, determined by which $\hbox{U}(1)$ is preserved. A standard choice is
to preserve the $T_3$ generator, which corresponds to the following $\mathbb{Z}_2$ transformation 
properties for the gravitini: $\Psi_M(-x^5) = i \tau_3 \gamma^{\dot 5} \Psi_M(x^5)$.
The full parity assignments are then listed in Table 1.

\vskip 10pt
\begin{table}[h]
\vbox{
$$\vbox{\offinterlineskip
\hrule height 1.1pt
\halign{&\vrule width 1.1pt#
&\strut\quad#\hfil\quad&
\vrule width 1.1pt#
&\strut\quad#\hfil\quad&
\vrule#
&\strut\quad#\hfil\quad&
\vrule#
&\strut\quad#\hfil\quad&
\vrule#
&\strut\quad#\hfil\quad&
\vrule#
&\strut\quad#\hfil\quad&
\vrule#
&\strut\quad#\hfil\quad&
\vrule#
&\strut\quad#\hfil\quad&
\vrule#
&\strut\quad#\hfil\quad&
\vrule width 1.1pt#
&\strut\quad#\hfil\quad&
\vrule#
&\strut\quad#\hfil\quad&
\vrule#
&\strut\quad#\hfil\quad&
\vrule#
&\strut\quad#\hfil\quad&
\vrule width 1.1pt#\cr
height3pt
&\omit&
&\omit&
&\omit&
&\omit&
&\omit&
&\omit&
&\omit&
&\omit&
&\omit&
&\omit&
&\omit&
&\omit&
&\omit&
\cr
&\hfil $\c\!$ field $\c\!\!$ &
&\hfil $\c e_M^A \c $&
&\hfil $\c \Psi_{M} \c$&
&\hfil $\c A_M \c$&
&\hfil $\c \vec{t} \c$&
&\hfil $\c v_{AB} \c$&
&\hfil \c $\vec{V}_M \c$&
&\hfil $\c \lambda \c$&
&\hfil $\c C \c$&
&\hfil $\c \vec{Y} \c $&
&\hfil $\c B_{MNP} \c$&
&\hfil $\c \rho \c$&
&\hfil $\c N \c$&
\cr
height3pt
&\omit&
&\omit&
&\omit&
&\omit&
&\omit&
&\omit&
&\omit&
&\omit&
&\omit&
&\omit&
&\omit&
&\omit&
&\omit&
\cr
\noalign{\hrule height 1.1pt} 
height3pt
&\omit&
&\omit&
&\omit&
&\omit&
&\omit&
&\omit&
&\omit&
&\omit&
&\omit&
&\omit&
&\omit&
&\omit&
&\omit&
\cr
&\hfil $\c\, + \c$ &
&\hfil $\c e_\mu^a, e_5^{\dot 5} \c$&
&\hfil $\c \psi^1_{\mu}, \psi^2_{5} \c$&
&\hfil $\c A_5 \c$&
&\hfil $\c t^{1,2} \c$&
&\hfil $\c v_{\alpha \dot 5} \c$&
&\hfil \c $V_\mu^3,V_5^{1,2} \c$&
&\hfil $\c \lambda^1 \c$&
&\hfil $\c C \c$&
&\hfil $\c Y^{1,2} \c$&
&\hfil $\c B_{\mu\nu\rho}\c$&
&\hfil $\c \rho^1\c$&
&\hfil $\c N \c$&
\cr
height3pt
&\omit&
&\omit&
&\omit&
&\omit&
&\omit&
&\omit&
&\omit&
&\omit&
&\omit&
&\omit&
&\omit&
&\omit&
&\omit&
\cr
\noalign{\hrule} 
height3pt
&\omit&
&\omit&
&\omit&
&\omit&
&\omit&
&\omit&
&\omit&
&\omit&
&\omit&
&\omit&
&\omit&
&\omit&
&\omit&
\cr
&\hfil $\c - \c$&
&\hfil $\c e_\mu^{\dot 5}, e_5^{a} \c$&
&\hfil $\c \psi^2_{\mu}, \psi^1_{5} \c$&
&\hfil $\c A_\mu \c$&
&\hfil $\c t^3 \c$&
&\hfil $\c v_{\alpha \beta} \c$&
&\hfil $\c V_\mu^{1,2},V_5^3 \c$&
&\hfil $\c \lambda^2 \c$&
&\hfil $\c \c$&
&\hfil $\c Y^3 \c$&
&\hfil $\c B_{\mu\nu 5}\c$&
&\hfil $\c \rho^2 \c$&
&\hfil $\c \c$&
\cr
height3pt
&\omit&
&\omit&
&\omit&
&\omit&
&\omit&
&\omit&
&\omit&
&\omit&
&\omit&
&\omit&
&\omit&
&\omit&
&\omit&
\cr
}
\hrule height 1.1pt}
$$
}
\noindent
\caption{\em Parity assignments for the bulk multiplets.}
\end{table}

At the fixed points, the even components of the 5D multiplets decompose into
multiplets of the supersymmetry preserved by the orbifold projection.
The even components associated to the 4D vielbein $e^\alpha_\mu$  fill up a 
so-called intermediate multiplet~\cite{Sohnius} given by 
$I=(e_\mu^\alpha,\psi_\mu^1; a_\mu, b_\alpha, t^1, t^2, \lambda^1,S)$ 
with the  identifications
\bea
S \a=\a C - \frac 12 e^5_{\dot 5}(\partial_5 t^3 - \bar \lambda^1 \psi^2_5
+ V_5^1 t^2 - V_5^2 t^1) \;, \\
a_\mu \a=\a -\frac 12 (V_\mu^3 - \frac 2{\sqrt{3}} F_{\mu 5} e^5_{\dot 5}
+ 4\, e_\mu^a \, v_{a \dot 5}) \;, \\
b_a \a=\a v_{a \dot 5} \raisebox{12pt}{} \;.
\eea
The vector $a_\mu$ gauges the R-symmetry~\cite{Sohnius}, and chiral multiplets are characterized 
by their chiral charge (or weight). The set of remaining even components forms a chiral multiplet 
$E^{\dot 5}_5=(e^{\dot 5}_5, \frac {2}{\sqrt{3}} A_5,\psi_5,V_5^1 - 4 t^2e_5^{\dot 5},
V_5^2 + 4 t^1e_5^{\dot 5})$ of weight $w=0$. However $E^{\dot 5}_5$ also transforms 
under 5D local translations and under the projected supersymmetry $\epsilon_2$. For instance,
$\delta \psi_5^2=\partial_5\epsilon_2+\dots$, which is not zero even at the boundary.
Because of this, $E^{\dot 5}_5$ cannot be used to write boundary Lagrangians. 
However the zero mode of $E^{\dot 5}_5$,
which is the only object that cannot be eliminated by choosing a suitable gauge for 5D
local supersymmetry and diffeomorphisms, remains as a chiral multiplet, the radion, of 4D supersymmetry.
Finally, all the even components of the compensator multiplet arrange into
a chiral multiplet $S_0=(Y^2, Y^1,\rho; {\rm Re}\,F_{S_0},{\rm Im}\,F_{S_0})$ of weight $w_0=2$ with:
\bea
{\rm Re}\,F_{S_0} \a=\a -2N + \hat {\cal D}_{\dot 5}Y^3 \;, \label{FS0}\\
{\rm Im}\,F_{S_0} \a=\a 2 W^{\dot 5} + 12(Y^2 t^1 - Y^1 t^2) \label{GS0} \;.
\eea

The fields localized at the fixed points can be either chiral multiplets, made of a complex scalar 
$\phi$, a chiral fermion $\chi$, and auxiliary fields: 
$\Phi = ({\rm Re}\,\phi,{\rm Im}\,\phi,\chi; {\rm Re}\,F_\Phi,{\rm Im}\,F_\Phi)$, or  
vector multiplets, consisting of a vector boson $B_\mu$, a Majorana fermion $\psi$, 
and an auxiliary field (in the Wess-Zumino gauge): $V=(B_\mu, \psi,D)$.

The Lagrangian of the complete theory describing interactions between bulk and 
brane multiplets has the general form
\bea
\Lag = \Lag_5 + \delta(x^5) \Lag_{4,0} + \delta(x^5-\pi)\Lag_{4,\pi} \;,
\eea
where $\Lag_5$ describes the dynamics of the minimal and compensator multiplets, whereas 
$\Lag_{4,0}$ and $\Lag_{4,\pi}$ describe the dynamics of the chiral and vector multiplets of 
the visible and hidden sectors and their interactions with the minimal and compensator multiplets. 

\subsection{Bulk Lagrangian}
\label{bulklag}

The bulk Lagrangian has been derived in~\cite{Zucker}. It is given by the
sum ${\cal L}_5 = {\cal L}_{\rm min} + {\cal L}_{\rm tens}$ of the Lagrangians for the 
gravity and compensator multiplets:
\bea
{\cal L}_{\rm min} \a=\a \Big[-32\vec{t}^2
-\frac{1}{\sqrt{3}}F_{AB}v^{AB} 
+ \bar{\Psi}_M\vec{\tau}\gamma^{MN}\Psi_N\vec{t}
- \frac{1}{6\sqrt{3}}\varepsilon^{MNPQR} A_M F_{NP} F_{QR} \nn \\
\a\;\a \hspace{4pt} +\, \frac{i}{8\sqrt{3}}\varepsilon^{MNPQR} \bar{\Psi}_M\gamma_N\Psi_P F_{QR} \Big] 
+ \Big[-4C -2i\bar{\lambda}\gamma^M\Psi_M\Big] \;, \\
{\cal L}_{\rm tens} \a=\a \Big[\!
-\frac{1}{4} Y {\cal R}(\widehat{\omega}) 
-\frac{i}{2} Y \bar{\Psi}_P\gamma^{PMN}{\cal D}_M\Psi_N 
- \frac{1}{6} Y \widehat{F}_{MN}\widehat{F}^{MN} 
- \frac{1}{4} Y^{-1} {\cal D}_M\vec{Y} {\cal D}^M\vec{Y} \nn \\
\a\;\a \hspace{7pt} 
+\, Y v_{AB}v^{AB} 
+ 20 Y \vec{t}^2 
+ Y^{-1} W_A W^A 
- Y^{-1} (N + 6\vec{t}\vec{Y})^2 
- Y \bar{\Psi}_M\vec{\tau}\gamma^{MN}\Psi_N \vec{t} \raisebox{15pt}{} \nn \\
\a\;\a \hspace{7pt} 
-\, \frac{i}{2} Y \bar{\Psi}_A\Psi_B v^{AB} 
- \frac{i}{4\sqrt{3}} Y \bar{\Psi}_M\gamma^{MNPQ}\Psi_N\widehat{F}_{PQ} 
- \frac{1}{24}Y^{-1}\varepsilon^{MNPQR}\vec{Y}\vec{G}_{M N} B_{PQR} \raisebox{22pt}{} \nn \\
\a\;\a \hspace{7pt} 
+\,\frac{1}{24} Y^{-3}\varepsilon^{MNPQR}\vec{Y}({\cal D}_M\vec{Y}\times{\cal D}_N\vec{Y})B_{PQR}
-\frac{1}{4}Y^{-1}\bar{\Psi}_A\vec{\tau}\gamma^{ABC}\Psi_B(\vec{Y}\times{\cal D}_C\vec{Y})\Big] \nn \\
\a\;\a +\, \Big[\mbox{terms involving $\rho$ but not $C$ or $\lambda$}\Big]
+ \Big[\,4 Y C + 2i Y \bar\lambda \gamma^A \Psi_A - 4 \bar{\lambda}\vec{\tau}\rho\vec{Y}\Big] 
\raisebox{18pt}{} \;.
\eea
The quantities $F_{MN}$ and $\vec G_{MN}$ are the field strengths of $A_M$ and $\vec V_M$ 
respectively, and $\widehat{F}_{MN} = F_{MN} + i\frac {\sqrt{3}}2\, \bar \Psi_M \Psi_N$. 
The covariant derivative ${\cal D}_M$ involves the ${\rm SU}(2)_R$ and super-Lorentz connections 
$\vec V_M$ and $\widehat{\omega}_{MAB} = \omega_{MAB} - \frac i2(\bar \Psi_A \gamma_M \Psi_B 
+ \bar \Psi_M \gamma_A \Psi_B - \bar \Psi_M \gamma_B \Psi_A)$, so that for instance 
${\cal D}_M\vec{Y} = \partial_M \vec Y + \vec V_M \times \vec Y$ and 
${\cal D}_M\Psi_N = D(\widehat{\omega})_M \Psi_N -\frac i2 \vec V_M \vec \tau\,\Psi_N$. 

In the situation that we shall consider in the following, matter does not couple to the Lagrange multipliers 
$C$ and $\lambda$. Their Lagrangian is thus given by the sum of the last brackets in ${\cal L}_{\rm min}$ 
and ${\cal L}_{\rm tens}$, and their equations of motion imply $Y=1$ and $\rho = 0$. All the terms in the 
second bracket in ${\cal L}_{\rm tens}$ are therefore irrelevant, and the Lagrangian simplifies to:
\bea
{\cal L}_5 \a=\a 
-\frac{1}{4} {\cal R}(\widehat{\omega}) 
-\frac{i}{2} \bar{\Psi}_P\gamma^{PMN}{\cal D}_M\Psi_N 
- \frac{1}{6} \widehat{F}_{MN}\widehat{F}^{MN} 
- \frac{1}{\sqrt{3}}\widehat{F}_{AB}v^{AB}
+ v_{AB}v^{AB} \nn \\
\a\;\a  
-\, 12 \vec{t}^2 
+ W_A W^A
-\,(N + 6\vec{t}\vec{Y})^2
- \frac{1}{4}{\cal D}_M\vec{Y}{\cal D}^M\vec{Y}
- \frac{1}{24}\varepsilon^{MNPQR}\vec{Y}\vec{G}_{MN} B_{PQR} \nn \\ 
\a\;\a 
-\, \frac{i}{4\sqrt{3}} \bar{\Psi}_M\gamma^{MNPQ}\Psi_N\widehat{F}_{PQ}
- \frac{1}{6\sqrt{3}}\varepsilon^{MNPQR} (A_M F_{NP} 
- \frac {3i}4 \bar{\Psi}_M\gamma_N\Psi_P )F_{QR} \nn \\
\a\;\a 
+\,\frac{1}{24} \varepsilon^{MNPQR}\vec{Y}({\cal D}_M\vec{Y}\times{\cal D}_N\vec{Y})B_{PQR} 
- \frac{1}{4}\bar{\Psi}_M\vec{\tau}\gamma^{MNP}\Psi_N(\vec{Y}\times{\cal D}_P\vec{Y})
\;.
\label{simplified}
\eea
The auxiliary scalar $\vec Y$ is forced to acquire a non-zero VEV, since it is constrained 
to satisfy $Y=1$. ${\rm SU}(2)_R$ is thus broken spontaneously, and a suitable gauge fixing 
is given by $\vec Y = (0,1,0)$. Notice that the VEV of $Y$ preserves the symmetry generated 
by $T_2$, while the orbifold preserves the one associated to $T_3$, so that no residual 
gauge symmetry survives the compactification. The bulk Lagrangian is then
\bea
{\cal L}_5 \a=\a 
-\frac{1}{4} {\cal R}(\widehat{\omega}) 
-\frac{i}{2} \bar{\Psi}_P\gamma^{PMN}{\cal D}^\prime_M\Psi_N 
- \frac{1}{6} \widehat{F}_{MN} \widehat{F}^{MN} 
- \frac{1}{\sqrt{3}}\widehat{F}_{AB}v^{AB}
+ v_{AB}v^{AB} \nn \\
\a\;\a  
-\, 12 \vec{t}^2 
+ W_A W^A
-\,(N + 6 t^2)^2
- \frac 14 \Big(V_{1A} V_1^{A} + V_{3A} V_3^{A}\Big)
- \frac{1}{12}\varepsilon^{MNPQR} \partial_M V_N^2 B_{PQR} \nn \\ 
\a\;\a 
-\, \frac{i}{4\sqrt{3}} \bar{\Psi}_M\gamma^{MNPQ}\Psi_N\widehat{F}_{PQ}
- \frac{1}{6\sqrt{3}}\varepsilon^{MNPQR} (A_M F_{NP} - \frac {3i}4 \bar{\Psi}_M\gamma_N\Psi_P )F_{QR} \;.
\label{L5ini}
\eea
Only $V_2^A$, corresponding to the unbroken $T_2$, appears now in the covariant derivative 
${\cal D}^\prime_M = D_M - \frac i2  V_M^2 \tau_2$. The terms involving the other vector 
auxiliary fields $V_1^A$ and $V_3^A$ have canceled against analogous interactions coming 
from the last term in (\ref{simplified}). Notice also that $V_2^A$ enters only linearly
in the Lagrangian, and after integrating by parts and using eq.~(\ref{WM}) all terms sum up to
$V_2^A W_A$.

\subsection{Boundary Lagrangians}
\label{boundarylag}

The Lagrangians $\Lag_{4,0}$ and $\Lag_{4,\pi}$ are constructed by using the tensor calculus of 
4D supergravity in the formalism of~\cite{Sohnius}, and consist of generic interactions involving 
the matter multiplets $\Phi$ and $V$, the gravitational intermediate multiplet $I$ and the compensator 
$S_0$. It is useful to briefly recall how 4D Lagrangians are constructed in the intermediate
multiplet formalism~\cite{Sohnius}. The presence of an extra set of auxiliary fields
leads to constraints on the chiral matter multiplets: with $n+1$ chiral multiplets
in the off-shell formulation, the constraints eliminate one combination of them, leading to
a $n$-dimensional K\"ahler manifold. Due to this fact, for a given physical on-shell Lagrangian
there is a family of off-shell Lagrangians which reduce to it. To make computations simpler
it is useful to write the off-shell Lagrangian in such a way that the constraint involves
just one multiplet with non-zero chiral weight, the compensator $\Xi$. Without loosing generality, 
but making contact with our 5D model (see below), we can take the compensator to have weight $w_\Xi=2$. 
The construction of a generic Lagrangian for $n$ chiral multiplets $\Phi_i$ is then straightforward.
Again, without loss of generality, we can choose all the $\Phi_i$ to have zero chiral weight.
(If $\Phi_i$ had weight $w_{\Phi_i}$, we could make it zero by a field redefinition  
$\Phi_i\to \Phi_i\, \Xi^{-w_i/2}$). Then any function $\Omega(\Phi_i,\Phi_i^\dagger)$ will be 
a vector superfield according to the tensor calculus of ref.~\cite{Sohnius}, where vector superfields 
have zero chiral weight. Moreover the expression $\Xi\, P(\Phi_i)$, for arbitrary $P$, is a chiral 
superfield of weight $2$, whose $F$ component has zero chiral weight and can be used to write a 
Lagrangian density. Then the 4D Lagrangian can be written as 
\be
\Lag_4^{\rm chi} = \Big[\Omega(\Phi,\Phi^\dagger)
\Big((\Xi\,\Xi^\dagger)^{r}-(1-3r)\Big)\Big]_D 
+\Big[P(\Phi)\,\Xi\Big]_F+\Big[P(\Phi)\,\Xi\Big]^\dagger_{F} \;.
\label{invariant}
\ee
Notice that the dependence of the $D$-term on $\Xi$ is to a large extent arbitrary, as long as 
it comes just through the vector multiplet $\Xi\,\Xi^\dagger$. Here we explicitly emphasized 
this fact by choosing an arbitrary exponent\footnote{In the superconformal approach \cite{supconf}, 
Weyl invariance constrains the $D$ term to be just $[\Omega\,(\Xi\,\Xi^\dagger)^{\frac 13}]_D$.} $r$. 
The equation of motion of the auxiliary scalar $S$ in the gravitational multiplet leads to a 
simple constraint for the scalar component $\xi$ of $\Xi$:
\be
\Big(\frac{1}{3}-r\Big)\Big(|\xi|^{2r} - 1\Big) = 0 \;.
\ee
Notice however that for $r = \frac 13$ the constraint disappears. For instance, after compactification
on $S_1/\mathbb{Z}_2$, in the absence of boundary terms, the effective off-shell Lagrangian for the 
light modes is
\be
\Lag^{\rm eff}=\Big[(T+T^\dagger)\Big({\sqrt{\Xi\,\Xi^\dagger}}+\frac 12 \Big)\Big]_D \;,
\label{effoff}
\ee
where the radion $T$ and the compensator $\Xi$ are just the zero modes of respectively the
$E_5^{\dot 5}$ and $S_0$ supermultiplets defined previously.

In writing the boundary action we should apply the above rules, with $S_0$ playing the role of 
the compensator $\Xi$. The freedom we have in the off-shell formulation can be exploited in order to
make the calculations simpler. In particular, the dependence on the bulk auxiliary
fields can be kept at a minimum by writing the boundary Lagrangian as
\be
\Lag_4^{\rm chi} = \Big[\Omega(\Phi,\Phi^\dagger)\Big(S_0S_0^\dagger\Big)^{\frac{1}{3}}\Big]_D 
+\Big[P(\Phi)S_0\Big]_F+\Big[P(\Phi) S_0\Big]^\dagger_{F} \;.
\label{semisimple}
\ee
This will become clear in the examples below. Actually, the basic situation that we will be mostly 
interested in is $\Omega(\Phi,\Phi^\dagger) = \Phi \Phi^\dagger$ and $P(\Phi)=0$. In this special 
case, it is convenient to choose $w_\Phi= \frac 23$ and write the boundary Lagrangian as
\be
\Lag_4^{\rm chi} = \Big[\Phi\Phi^\dagger\Big]_D \;.
\label{simple}
\ee
We will see that with this specific off-shell formulation several auxiliary field do not couple 
to matter and can be integrated out at the classical level to yield a formulation which is still 
off-shell enough to correctly describe interactions and reproduce supersymmetric cancellations at 
the quantum level. 

Let us now work out the component expressions of the boundary actions describing the interaction 
of chiral and vector multiplets $\Phi$ and $V$ with the intermediate multiplet $I$. To simplify 
the formulae, we will only write the relevant pieces of the Lagrangians, neglecting all interaction
terms involving fermions. Consider first the Lagrangian (\ref{simple}) for a chiral multiplet 
$\Phi$ with generic chiral weight $w$. Defining the complex auxiliary field $t=t^2 + i t^1$, its 
explicit component expression reads: 
\bea
\Lag_4^{\rm chi} \a=\a |{\cal D}_\mu \phi|^2 + i \bar\chi \gamma^\mu {\cal D}_\mu \chi 
+ |F_\Phi - 4 \phi\,t^*|^2 + \frac w4\, |\phi|^2 \big({\cal R} 
+ 2i \bar \psi_\mu^1 \gamma^{\mu \nu \rho} D_\nu \psi_\rho^1\Big) \nn \\ 
\a\;\a +\, 6 (w - \frac 23) |\phi|^2 
\Big[b_\mu^2 - 2 S - 8 |t|^2 \Big] + \cdots \;,
\label{L4chi}
\eea
where the chiral covariant derivatives are given by
\bea
{\cal D}_\mu \phi \a=\a \partial_\mu \phi + i w\,\Big(a_\mu + \frac 2w\, b_\mu\Big) 
\phi \label{covchi1} \;, \\
{\cal D}_\mu \chi \a=
\a D_\mu \chi - i (1 - w)\,\Big(a_\mu + \frac{1}{1-w}\, b_\mu\Big) \chi \;.
\label{covchi2}
\eea
The Lagrangian $\Lag_4^{\rm vec}$ for a vector multiplet $V$ has already  been worked out 
in~\cite{Zucker}, and we therefore quote only the result:
\bea
\Lag_4^{\rm vec} \a=\a - \frac 14 G_{\mu\nu}^2 + i \bar \psi \gamma^\mu {\cal D}_\mu \psi
+ \frac 14 D^2 + \cdots \;,
\label{L4vec}
\eea
where 
\bea
{\cal D}_\mu \psi = D_\mu \psi - \gamma^{\dot 5} \Big(a_\mu + 3\, b_\mu\Big) \psi \;.
\label{covvec}
\eea
As anticipated, a substantial simplification occurs when the chiral multiplet has weight 
$w=\frac 23$. In this case, the second line in (\ref{L4chi}) drops out and there is therefore 
no tadpole for $S$, as already assumed in previous subsection. Moreover, the same combination 
of auxiliary fields $a_\mu + 3 b_\mu = - \frac 12\, (V_\mu^3 - \frac 2{\sqrt{3}} F_{\mu  5} e_5^{\dot 5} 
- 2\,e_\mu^\alpha  v_{\alpha \dot 5})$ is left in all the covariant derivatives (\ref{covchi1}), 
(\ref{covchi2}) and (\ref{covvec}).

There is actually a simple generalization of the basic situation 
$\Omega(\Phi,\Phi^\dagger) = \Phi \Phi^\dagger$ that we would like to consider. 
It consists in adding a real constant kinetic function $\Omega=-3L$. 
The simplest way to construct the additional terms in the off-shell boundary Lagrangians 
is to use now eq.~(\ref{semisimple}). In this case, a non-trivial dependence on the 
compensator auxiliary fields $N$ and $W_{\dot 5}$ will appear. The possibility of having 
$\Omega=-3 L$ corresponds to adding localized kinetic terms for the bulk supergravity fields, 
and is required to construct kinetic functions of the form (\ref{lowest0}) and (\ref{lowestpi}).
The component expansion of the corresponding action is easily found to be:
\be
\Lag_4^{\rm loc} = - \frac L2 \Big[{\cal R} + 2 i \bar \psi_\mu^1 \gamma^{\mu\nu\rho} D_\nu \psi_\rho^1 
+ \frac 83 (a_\mu + 3 b_\mu)^2 + \frac 83 (N + 6 t^2 - \frac 12 V_{\dot 5}^1)^2 
+ \frac 83 W_{\dot 5}^2 + \cdots \Big] \;.
\label{localized}
\ee
As in the minimal situation, the auxiliary fields $a_\mu$ and $b_\mu$ appear only in the 
universal combination $a_\mu + 3\,b_\mu$. Moreover, the additional dependence on the auxiliary 
fields $N$, $t^2$, $V_{\dot 5}^2$ and $W^{\dot 5}$ occurs only in the two combinations 
$N + 6 t^2 - \frac 12 V_{\dot 5}^1$ and $W^{\dot 5}$. This will be important in next section, 
in which most of these fields will be integrated out.

\subsection{Partially off-shell formulation}
\label{partoff}

The only auxiliary fields that are influenced by the boundary are $V_\mu^3$, 
$v_{\alpha \dot 5}$, $t_1$ and $t_2$, as well as $N$, $V_{\dot 5}^1$ and 
$B_{MNP}$ if constant kinetic functions are included. All the other 
auxiliary fields can then be integrated out just by using (\ref{L5ini}), to 
give a partially off-shell formulation which is still powerful enough to 
correctly describe all bulk-to-boundary interactions. The equations of motion 
of the fields $t_3$, $v_{\alpha\beta}$, $V_1^\alpha$, $V_2^A$ and $V_3^{\dot 5}$ 
are trivial and imply 
$t_3 = 0$, $v_{\alpha\beta} = \frac 1{2\sqrt 3} \hat F_{\alpha\beta}$, 
$V_1^\alpha = 0$, $W^A = 0$ and $V_3^{\dot 5} = 0$. Since $W^{\dot 5} = 0$, 
the dependence on $B_{MNP}$ coming from the boundary Lagrangian (\ref{localized}) 
trivializes, and its equation of motion can be derived from the bulk Lagrangian 
(\ref{L5ini}) as well. It leads to the condition that the field strength of $V_2^M$ 
vanish: $\partial_M V_{2N} - \partial_N V_{2M} = 0$. This implies that the 
connection $V_2$ is closed. Since spacetime is in this case not simply connected, 
$V_{2}$ is not necessarily exact and can have a physical effect, parametrized by 
the gauge-invariant quantity $\epsilon = \int dx^5\,V_2^5(x^5)$. 
This is a Wilson line for the unbroken $\hbox{U}(1)_{T_2}$, and it is equivalent to 
Scherk--Schwarz supersymmetry breaking with twist $\epsilon$~\cite{vonGersdorff1}. 
In section \ref{calcolo} we will explain this in more detail.

The  auxiliary fields $N$ and $V_{\dot 5}^1$ 
appear both in the bulk Lagrangian (\ref{L5ini}) and in the boundary Lagrangian 
(\ref{localized}), but their effect is nevertheless trivial. This is most easily 
seen by first substituting them with the two new combinations 
$N_\pm = N + 6 t^2 \pm \frac 12 V_{\dot 5}^1$. These appear in the bulk Lagrangian 
(\ref{L5ini}) only through a term proportional to $N_+ N_-$, whereas in the boundary 
Lagrangian (\ref{localized}) only a term proportional to $N_-^2$ appears. The equation 
of motion of $N_-$ fixes therefore the value of $N_+$, but that of $N_+$ implies  
$N_- = 0$, so that all the dependence on $N_\pm$ has finally no effect. This is 
perfectly analogous to what happens in 4D no-scale models, where the equation of 
motion of $F_T$ enforces the condition $F_{S_0}=0$.

To proceed further, it is convenient to redefine the remaining auxiliary fields in 
such a way to disentangle those combinations of them which do not couple to matter 
and integrate them out. This is most conveniently done by defining the following new 
vector and scalar auxiliary fields: 
\bea
V_\alpha \a=\a e_\alpha^M V_M^3 -\frac 2{\sqrt{3}} e_\alpha^M F_{M  5}e^5_{\dot 5} 
- 2 v_{\alpha \dot 5} \;, \label{redefinition} 
\eea
Notice that since at the boundary we have $e_\mu^{\dot 5}=e_5^\alpha=0$, the vector that couples 
to the boundary is $V_\mu \equiv e_\mu^\alpha V_\alpha = -2(a_\mu + 3 b_\mu)$. Thanks to the 
above redefinitions, $v_{\alpha \dot 5}$ no longer couples to matter and can be integrated out 
through its equation of motion $v_{\alpha\dot 5} = \frac 1{2\sqrt{3}}\hat F_{\alpha\dot 5}$. 
Similarly, the equation of motion of $V_{\dot 5}^3$ now trivially implies $V_{\dot 5}^3 = 0$.
After a straightforward computation, splitting the covariant derivatives and factoring out the 
volume element $e={\rm det}({e_M^A})$ explicitly\footnote{$e_{\dot 5}^5\,\delta(x^5)$ is the 
scalar $\delta$-function density}, we finally find
\bea
e^{-1}\! \Lag \a=\a 
\frac 16 \Omega(x^5) \Big[{\cal R} + 2i \bar \Psi_M \gamma^{MNP}\! D_N \Psi_P 
+ \frac 23 V_\alpha V^\alpha \Big] - 12 |t|^2 \nn \\
\a\;\a +\, \Omega_{\phi\phi^*}(x^5) \Big[|\partial_\mu \phi|^2 
+ i \bar \chi D\!\!\!\!/\, \chi + |F_\Phi - 4 \phi t^*|^2\Big] 
+ e_{\dot 5}^5\,\delta(x^5) \Big[\!-\! \frac 14 G_{\mu\nu}^2 
+ i \bar \psi D\!\!\!\!/\, \psi + \frac 14 D^2\Big] \nn \\
\a\;\a -\, \frac 14 F_{\alpha\beta}^2 
+ \frac 13 \Big(J_\alpha^{\rm mat}(x^5) - \sqrt{3} F_{\alpha \dot 5} \Big) V^\alpha 
+ \cdots \;.
\label{Ltilde}
\eea
In this expression, $\Omega(x^5)$ is a generalized kinetic function defined as
\bea
\Omega(x^5) \a=\a - \frac 32 + \Big(-3L + |\phi|^2\Big) e_{\dot 5}^5\,\delta(x^5) \;.
\label{Omegax5}
\eea
It is understood that the localized part of $\Omega(x^5)$ multiplies only 
the restrictions of the kinetic terms to the boundary. 
Similarly, $J_\mu^{\rm mat}(x^5)= J_\mu^{\rm chi}(x^5) + J_\mu^{\rm vec}(x^5)$ is a 
generalized matter $R$-symmetry current, defined by\footnote{The $R$-charge of 
$\phi$, $\chi$ and $\psi$ are equal respectively to $\frac 23$, $-\frac 13$ and 
$-1$, but for convenience we take out an overall factor of $\frac 23$ in the 
definition of the current.}:
\bea
J_\mu^{\rm chi}(x^5) \a=\a i(\Omega_\phi (x^5)\partial_\mu \phi - \mbox{c.c.}) 
- \frac i2 \Omega_{\phi\phi^*} (x^5) \bar \chi \gamma_\mu \gamma^{\dot 5} \chi 
+ \cdots \;, \label{Jchix5} \\
J_\mu^{\rm vec}(x^5) \a=\a \frac {3i}2 e_{\dot 5}^5\, \delta(x^5) 
\bar \psi \gamma_\mu \gamma^{\dot 5} \psi \;.
\label{Jvecx5}
\eea
Finally, the dots denote boundary terms describing the standard 4D supergravity 
interactions of the gravitino with matter, the only truly novel interaction 
between bulk and brane being those with $V_\mu$.

The field $V_\mu$ is the analog of the vector auxiliary field $b_\mu$ of Poincar\'e 
supergravity~\cite{supconf,WessBagger},
but it mixes with the graviphoton $A_M$, and is therefore no longer an ordinary auxiliary field. 
The graviphoton has also changed its dynamics: the KK mass term $\frac 12 F_{\mu 5}^2$ has 
disappeared. 5D covariance is not manifest because of the non-covariant field redefinition 
of eq.~(\ref{redefinition}); by integrating out $V_\alpha$, however, we would recover the 
fully covariant graviphoton kinetic term. \footnote{The field $t$ is similar to the auxiliary 
field $M$ of Poincar\'e supergravity ~\cite{supconf,WessBagger}, but it does not coincide 
with it.}

The Lagrangian (\ref{Ltilde}) that we find is perfectly analogous to the one found by Mirabelli 
and Peskin~\cite{Mirabelli} in the case of a 4D chiral multiplet interacting with a 5D vector 
multiplet. There the role of $V_\mu^3$  and $A_\mu$ is  played respectively by $X^3$, $T_3$-singlet 
component of the auxiliary field $\vec X$, and by $\Sigma$, the extra physical scalar of 
the 5D vector multiplet. The boundary couples only to the combination $X=X^3-\partial_5 \Sigma$, 
which plays the role of $V_\mu$. The propagation of $X$, $\Sigma$ and their interaction with 
the boundary is described by
\be
\Lag_{X,\Sigma} = \frac{1}{2}\partial_\mu\Sigma \partial^\mu \Sigma 
+ X\partial_5\Sigma - \frac{1}{2}X^2 +\delta(x^5)X |\phi|^2 \;.
\label{globalcase}
\ee
Notice that, like in our case, the auxiliary field $\Sigma$ propagates in the 5th dimension 
only via its mixing to $X$.

From eq.~(\ref{Ltilde}) one would normally go ahead and eliminate the remaining auxiliary 
fields to write the physical Lagrangian. For $F_\Phi$ and $t$ this can be trivially done. On the 
other hand, $V_\mu$ has sources proportional to $\delta(x^5)$ so that after solving its 
equation of motion the physical Lagrangian contains seemingly ambiguous expressions involving 
powers of $\delta(x^5)$. Indeed, since the kinetic term of $V_\mu$ has a coefficient given by 
eq.~(\ref{Omegax5}), the effective Lagrangian, proportional to $1/\Omega(x^5)$, will formally 
involve infinite powers of $\delta(x^5)$. This should be compared to the global case of 
ref.~\cite{Mirabelli}, see eq.~(\ref{globalcase}), where one has ``just'' to deal with 
$\delta^2(x^5)$. Now, the presence of tree level UV divergences is a normal fact in theories 
with fixed points: the momentum in the orbifolded directions is not conserved so that the 
momentum on the external lines does not fix the virtual momenta even at tree level. For 
propagating fields in $n$ extra dimensions the sum over the transverse momentum $p_T$ gives 
rise to an amplitude
\be
\int \frac{d^n p_T}{p^2+p_T^2}
\ee
which leads to UV divergences when $n\geq 2$. For an auxiliary field, the propagator is 
just $1$, so the UV divergences appear already with $n=1$. However, in the case at hand, 
these UV divergences are a spurious effect of integrating out an incomplete supermultiplet. 
In physical quantities they will never appear. Physically we should also account for the 
propagation of the graviphoton $A_\mu$ (or of $\Sigma$ in the global case). Notice that 
$A_5$ plays no role as we can choose the gauge $\partial_5 A_5=0$ where it has no local 
5D degrees of freedom. The mixed $A_\mu, V_\mu$ kinetic matrix has then the form
\bea
K_{A,V}=\left(\matrix {p^2\eta_{\mu\nu}-p_\mu p_\nu \a
\displaystyle{\frac{1}{2\sqrt 3}}\, p_5\,\eta_{\mu\nu} \smallskip \cr 
\displaystyle{\frac{1}{2\sqrt 3}}\, p_5\,\eta_{\mu\nu} \a
\displaystyle{\frac{1}{3}}\,\eta_{\mu\nu}\cr}\right) \;.
\label{KVA}
\eea     
The propagator of $A_\mu$ and $V_\mu$ is obtained by inverting this matrix. Since the $AA$ entry 
does not involve any $p_5^2$, the $\langle V_\mu V_\nu\rangle$ propagator scales like $p^2/p_5^2$,
and the exchange of $V_\mu$ between boundary localized sources does not lead to any UV divergences. 

One example of a physical object that is calculated by integrating out the auxiliary KK modes is the 
low-energy two-derivative effective Lagrangian after compactification. In order to compute it, we will 
pick the zero modes of the physical fields $e_\mu^\alpha(x,x^5)\equiv e_\mu^\alpha(x)$ and similarly 
for $\psi_\mu^1$, $\psi_5^2$ and $A_5$ without changing notation. On the other hand, we set 
$e_5^\alpha=e_\mu\hspace{-5pt}\raisebox{3pt}{$\scriptstyle{\dot 5}$}\,\equiv 0$, so that indices 
are raised and lowered according to 4D rules.
Finally we define the radion field by $e_5^{\dot 5}(x,x^5)\equiv R(x)$ and normalize the radion 
supermultiplet\footnote{The relative coefficients of the real and imaginary parts of $T$ agree with 
Luty and Sundrum (LS)~\cite{Luty}, after noticing that our $A_5$ equals $\frac 1{\sqrt{2}} A_5^{LS}$ due 
to our different normalization of the supergravity kinetic terms. Notice also our different overall 
normalizations: $T_{LS}=3 T$.} as $T/\pi=(R + i\frac {2}{\sqrt{3}}A_5, \psi_5^2)$. The graviphoton 
$A_\mu$ does not have zero modes and it is conveniently integrated out by working in the gauge 
$\partial_5 A_5=0$, where only the physical zero mode of $A_5$ is turned on. The $\partial_\mu A_5/R$ 
piece in $F_{\mu \dot 5}$ corresponds to the radion contribution to the generalized $R$-symmetry current:
\be
J_\mu^{\rm rad}(x^5) = -\frac {3 i}{2(T + T^\dagger)}(\partial_\mu T - {\rm c.c.}) \;.
\label{Jradx5}
\ee
This reconstructs the total $R$-current $J_\mu(x^5) = J_\mu^{\rm mat}(x^5) + J_\mu^{\rm rad}(x^5)$
in the last term of (\ref{Ltilde}). The graviphoton $A_\mu$ can now be integrated out at the classical
level. Neglecting the $F_{\mu\nu}^2$ term, which only affects higher-derivative terms in the low-energy 
action, the $A_\mu$ equation of motion amounts to the constraint
\bea
\partial_5 V_\mu = 0 \;,
\eea
saying that only the zero mode of $V_\mu$ survives. As $V_\mu$ is constant, to obtain the low-energy 
effective action we just need to integrate eq.~(\ref{Ltilde}) over $x^5$; the result is
\bea
\Lag^{\rm eff} \a=\a
\frac 16 \Omega \Big[{\cal R} + 2 i \bar \psi_\mu^1 \gamma^{\mu\nu\rho} D_\nu \psi_\rho^1 
+ \frac 23 V_\mu^2 \Big] + \frac 13 J_\mu V^\mu \nn \\
\a\;\a +\, \Omega_{\phi\phi^*} \Big[|\partial_\mu \phi|^2 + i \bar \chi D \!\!\!\!/\, \chi \Big] 
+ \Big[- \frac 14 G_{\mu\nu}^2 + i \bar \psi D \!\!\!\!/\, \psi\Big] + \cdots \;. 
\label{Lefftilde}
\eea
In this expression, the 4D quantities $\Omega$ and $J$ are obtained by integrating the 
corresponding generalized 5D quantities $\Omega(x^5)$ and $J(x^5)$, defined 
by eq.~(\ref{Omegax5}) and the sum of (\ref{Jchix5}), (\ref{Jvecx5}), (\ref{Jradx5}), 
over the internal space. Denoting the former with $X$ and the latter with $X(x^5)$, the 
precise relation is $X = \int_{-\pi}^{\pi} dx^5\, e_5^{\dot 5}\,X(x^5)$. The kinetic 
function is found to be
\bea
\Omega \a=\a - \frac 32 (T + T^\dagger) - 3 L + |\phi|^2 \;.
\label{Omega}
\eea
and the total $R$-symmetry current of the light fields 
$J_\mu = J_\mu^{\rm chi}+J_\mu^{\rm vec} + J_\mu^{\rm rad}$ is correctly reproduced 
with
\bea
J_\mu^{\rm chi} \a=\a 
i(\Omega_\phi \partial_\mu \phi - \mbox{c.c.}) 
- \frac i2 \Omega_{\phi \phi^*} \bar\chi \gamma_\mu \gamma^{\dot 5} \chi + \cdots \;, 
\label{Jchi} \\
J_\mu^{\rm vec} \a=\a 
\frac {3i}2 \bar\psi \gamma_\mu \gamma^{\dot 5} \psi \;, 
\label{Jvec} \\
J_\mu^{\rm rad} \a=\a 
i(\Omega_T \partial_\mu T - \mbox{c.c.}) \raisebox{15pt}{} \;.
\label{Jrad}
\eea
In the Lagrangian (\ref{Lefftilde}) $V_\mu$ is identified with the standard vector auxiliary field 
of 4D supergravity. It is easy to check, using for instance the formulae in~\cite{WessBagger}, that 
all coefficients in the above equations are correct. 

\subsection{On-shell formulation}

In this section we will compute the on-shell Lagrangian. We do that mainly to make contact with
the standard approach followed by Mirabelli and Peskin~\cite{Mirabelli}. We believe that our 
discussion completes or even corrects previous treatments of this issue in the supergravity case 
\cite{Gherghetta,Antoniadis}. 

Let us start from eq.~(\ref{Ltilde}). The most natural way to proceed is to complete the quadratic 
form depending on the auxiliary field $V_\alpha$ through a shift. This is achieved by defining the 
new auxiliary field
\be
\tilde V_\alpha =  V_\alpha + \frac{3}{2\Omega(x^5)} \Big[J_\alpha^{\rm mat}(x^5)
- \sqrt{3} F_{\alpha \dot 5}  \Big] \;,
\label{V}
\ee
where $\Omega(x^5)$ has been defined in eq.~(\ref{Omegax5}) and 
$J_\mu^{\rm mat}(x^5) = J_\mu^{\rm chi}(x^5) + J_\mu^{\rm vec}(x^5)$ in 
(\ref{Jchix5}) and (\ref{Jvecx5}).
Notice that we are working with the ill-defined distribution $1/\Omega(x^5)$. In what follows, one 
could think of $\delta(x^5)$ as being regulated. In the end, as evident from the discussion in the 
previous section, the regulation will not matter in the computation of physical quantities. After 
some straightforward algebra, and integrating out the trivial auxiliary fields $Q$, $F_\Phi$ and $D$, 
the Lagrangian can be rewritten as
\bea
e^{-1} \Lag \a=\a 
\frac 16 \Omega(x^5) \Big[{\cal R} + 2i \bar \Psi_M \gamma^{MNP} D_N \Psi_P 
+ \frac 23 {\tilde V}_\alpha^2 \Big] \nn \\
\a\;\a +\, \Omega_{\phi\phi^*}(x^5) \Big[|\partial_\mu \phi|^2 
+ i \bar \chi D\!\!\!\!/\, \chi \Big] 
+ e_{\dot 5}^5\, \delta(x^5) 
\Big[\!-\! \frac 14 G_{\mu\nu}^2 + i \bar \psi D\!\!\!\!/\, \psi\Big] \nn \\
\a\;\a - \frac 14 F_{\alpha\beta}^2 - \frac{3}{4 \Omega(x^5)} 
\Big[F_{\alpha \dot 5} - \frac 1{\sqrt{3}} J_\alpha^{\rm mat} (x^5)\Big]^2 
+ \cdots \;.
\label{L}
\eea
Notice that we have not truly integrated out $ V_\alpha$, but just rewritten the Lagrangian in terms 
of the classically irrelevant field $\tilde V_\alpha$. The reason for keeping $\tilde V_\alpha$
is that its kinetic term is field-dependent and gives rise to a Jacobian at the quantum level.
The above Lagrangian differs from the one advocated in~\cite{Gherghetta}; in particular, the 
interaction of the chiral multiplet with the graviphoton involves a non-trivial denominator with 
$\delta$-functions, which is crucial to correctly reproduce the quartic coupling of the effective 
4D theory (and of course to obtain the supersymmetric cancellations at the quantum level). 
More insight in these couplings can be obtained be expanding the perfect square to isolate the 
complete bulk kinetic term of the graviphoton. At leading order in a power series expansion in the 
scalar fields, one finds that the exceeding $F_{\alpha \dot 5}^2 |\phi|^2 e_{\dot 5}^5\,\delta(x^5)$ 
term just provides the correct scalar seagull correction to the coupling $J^\alpha(x^5) F_{\alpha \dot 5}$ 
to turn it into a minimal coupling through a covariant derivative, so that the $R$-symmetry appears 
to be gauged by $F_{\alpha \dot 5}$.

\medskip

We now show once more that the correct low-energy effective 4D theory is obtained when integrating 
out the heavy KK modes. Again, since we take $e_5^\alpha=e_\mu^{\dot 5}\equiv 0$ we can restore 
the curved indices to integrate out the  massive modes of the graviphoton. As before we work in the 
gauge $\partial_5 A_5=0$, and the $\partial_\mu A_5/R$ piece in $F_{\mu \dot 5}$ again corresponds 
to the radion contribution to the generalized $R$-symmetry current. Neglecting as before terms with 
4D spacetime derivatives with respect to $x^5$-derivatives in the low energy limit, and defining
the total generalized $R$-symmetry current $J_\mu(x^5) = J_\mu^{\rm mat}(x^5) + J_\mu^{\rm rad}(x^5)$ 
with $J_\mu^{\rm rad}(x^5)$ given by (\ref{Jradx5}), the Lagrangian for the heavy field $A_\mu$ can 
be written as
\be
{\cal L}_{A} \simeq - \frac{3}{4 \Omega(x^5)} 
\Big[\partial_5 A_\mu - \frac{1}{\sqrt{3}} J_\mu(x^5)\Big]^2 \;.
\ee
The corresponding equation of motion yields
\be
\partial_5 A_\mu = \frac{1}{\sqrt{3}} \Big[J_\mu(x^5) 
- \frac {\Omega(x^5)}{\Omega}J_\mu \Big] \;,
\ee
where the 4D kinetic function $\Omega$ and $R$-current $J_\mu$ arise again as integrals of
their 5D generalizations $\Omega(x^5)$ and $J_\mu(x^5)$. Plugging this expression back into the 
Lagrangian, discarding the auxiliary field and integrating over $x^5$, one finds finally the standard 
on-shell expression for a 4D chiral no-scale supergravity model with kinetic function $\Omega$ and 
vanishing superpotential: 
\bea
\Lag^{\rm eff} \a=\a
\frac 16 \Omega \Big[{\cal R} + 2 i \bar \psi_\mu^1 \gamma^{\mu\nu\rho} D_\nu \psi_\rho^1 \Big] 
- \frac{1}{4 \Omega} J_\mu^2  \nn \\
\a\;\a +\, \Omega_{\phi\phi^*} \Big[|\partial_\mu \phi|^2 + \bar \chi D\!\!\!\!/\, \chi \Big]
+ \Big[- \frac 14 G_{\mu\nu}^2 + i \bar \psi D\!\!\!\!/\, \psi \Big] + \cdots \;.
\label{Leff}
\eea

\medskip

\setcounter{equation}{0}
\section{Loop corrections to matter operators}
\label{test}

Before starting the computation outlined in the introduction, we shall verify in this section 
that the one-loop corrections to operators involving scalar fields and no derivatives correctly 
cancel as a consequence of the supersymmetry surviving the orbifold projection. In order to do 
that, we need to discuss the structure of the propagators of 5D fields. For the gauge field 
$A_M$ and the graviton $h_{MN}$ defined by expanding the metric around the flat background as 
$g_{MN} = \eta_{MN} + 2\sqrt{2} h_{MN}$, one can proceed along the lines of~\cite{Contino}.
For the gravitino, that we can now describe with an ordinary Dirac spinor\footnote{The kinetic 
term has then an additional factor of 2.} $\Psi_M = (\chi^1_M, \bar \chi^2_M)^T$, we refer 
instead to~\cite{Fung,VanNieuwenhuizen}. The mode expansions are standard and lead to towers 
of KK states with masses $m_n = n/R$. As usual it is convenient to use the doubling trick and 
run $n$ from $-\infty$ to $+\infty$, including $n=0$ with the same weight. For the gravitino, 
we use Dirac modes $\Psi_{n}^M = (\chi^{1M}_{n},\bar \chi^{2M}_{n})^T$. For simplicity we restrict 
to the basic case of a simple quadratic kinetic function and set $L=0$.

\subsection{On-shell formulation}

We consider first the completely on-shell formulation (\ref{L}), and focus 
on the simplest example of the class of operators we want to study: the scalar 
two-point function at zero momentum, i.e.\ the correction to the scalar mass.
The relevant interactions on the brane are easily obtained by expanding all interactions 
in (\ref{L}) to quadratic order and recalling the usual supersymmetric interaction between the
gravitino and the improved supersymmetric current of the chiral multiplet. To switch to
the new Dirac notation for the gravitino, we use the projectors 
$P_{L,R} = \frac 12 (1 \!\pm\! i\gamma^{\dot 5})$.
The terms that are relevant at zero momentum are given by:
\bea
\Lag^{\rm int} \a\simeq\a \delta(x^5)\,e_4\,\Big[\frac 13 |\phi|^2 \Big(\,\frac 12 {\cal R}_4
+ 2 i \bar \Psi_\mu \gamma^{\mu\nu\rho} P_L \partial_\nu \Psi_\rho 
+ \frac 13 \tilde{V}_\mu^2 + F_{\mu \dot 5}^2 \Big) \nn \\
\a\;\a \hspace{41pt} +\, \frac 13 \Big(\sqrt{2} \,\phi^* \bar \chi 
\gamma^{\mu \nu}  P_L \partial_\mu \Psi_\nu - i\sqrt{3} F_{\mu \dot 5} 
\phi^* \partial^\mu \phi - \mbox{c.c.}\Big)  \nn \\
\a\;\a \hspace{41pt} +\, \frac 16 |\phi^* \partial^\mu \phi - \mbox{c.c.}|^2 
e_{\dot 5}^5\, \delta(0) \Big] \;,
\label{Sint1}
\eea
where:
\be
e_4 {\cal R}_4 = 2\sqrt{2} \Big[\partial_\mu \partial_\nu h^{\mu\nu} - \partial^2 h\Big]
+ 2\Big[h \partial^2 h - h_{\mu \nu} \partial^2h^{\mu\nu} 
- 2 h^{\mu \nu} \partial_\mu \partial_\nu h \nn \\
+ 2 h^{\mu \nu} \partial_\nu \partial_\rho h^{\rho}_{\;\mu}\Big] \;.
\ee
As advertised in last section, the couplings between scalars and graviphotons reconstruct 
a minimal coupling with a covariant derivative given by ${\cal D}_\mu = \partial_\mu 
+ \frac i{\sqrt{3}} F_{\mu \dot 5}$.

To derive the propagators of the bulk fields, one has to chose a gauge. Unitary gauges \cite{Contino} have 
the advantage of explicitly disentangling physical and unphysical modes for massive KK modes, 
which will therefore have the propagators of standard massive particles. However, in general 
they do not fully fix the gauge for the zero modes, which must be separately specified. 
Moreover, the latter remain entangled in any gauge. For these reasons, it is more convenient 
to use covariant gauges which treat massless and massive modes on equal footing. For the 
graviphoton, the above problem does not exist, because $A_\mu$ does not have zero modes, and 
for later convenience we will thus choose the unitary gauge $\partial_5 A_5=0$. The propagators 
of the various modes are then given by
\be
\langle A_\mu A_\nu \rangle_n = -\Big[\eta_{\mu\nu} - \frac {p_\mu p_\nu}{m_n^2}\Big] 
\frac{i}{p^2-m_n^2} \;,\;\; 
\langle A_5 A_5 \rangle_0 = \frac{i}{p^2} \;. 
\label{propA}
\ee
For the graviton and the gravitino, we shall instead choose
the harmonic gauges (called de Donder in the case of the graviton)
and add to the 5D Lagrangian the gauge fixing terms
\bea
{\cal L}_{h}^{\rm GF} \a=\a - \Big[\partial_M (h^{MN}\! - \frac 12 \eta^{MN} h)\Big]^2 \;,\\
{\cal L}_{\Psi}^{\rm GF} \a=\a \frac i2 \bar \Psi_M \gamma^M \partial\!\!\!/ \gamma^N \Psi_N \;.
\eea
In these gauges, the propagators have a structure that is reminiscent 
of the 5D origin of the fields, and can be deduced by repeating the analysis of~\cite{Fung} 
on the orbifold after decomposing the fields in KK modes. For the 4D components, relevant 
to our computation, one finds:
\bea
\langle h_{\mu\nu} h_{\alpha\beta}\rangle_n \a=\a 
\frac 12 \Big[\eta_{\mu\alpha} \eta_{\nu\beta} 
+ \eta_{\mu\beta} \eta_{\nu\alpha}
- \frac 23 \eta_{\mu\nu} \eta_{\alpha\beta}\Big]\,\frac{i}{p^2-m_n^2} 
\raisebox{20pt}{} \label{proph} \;, \\
\langle \Psi_\mu \bar \Psi_\nu\rangle_n \a=\a 
\frac 16 \Big[- \gamma_\nu (p\!\!\!/ - m_n)\gamma_\mu 
+ \Big(\eta_{\mu\nu} - 2\, \frac {p_\mu p_\nu}{p^2-m_n^2}\Big) (p\!\!\!/ + m_n)\Big]
\frac i{p^2-m_n^2} 
\label{proppsi} \;.
\eea
Finally, the propagator of the auxiliary field $\tilde V_\mu$ is given in the 
same notation by
\bea
\langle \tilde V_\mu \tilde V_\nu \rangle_n \a=\a 
-3i\,\eta_{\mu\nu} \raisebox{16pt}{} 
\label{propV} \;.
\eea
Notice that in our computation at vanishing external momentum, the longitudinal pieces of the
propagators are actually irrelevant, because the couplings in (\ref{Sint1}) feel only the 
transverse polarizations and each diagram is gauge-independent on its own.

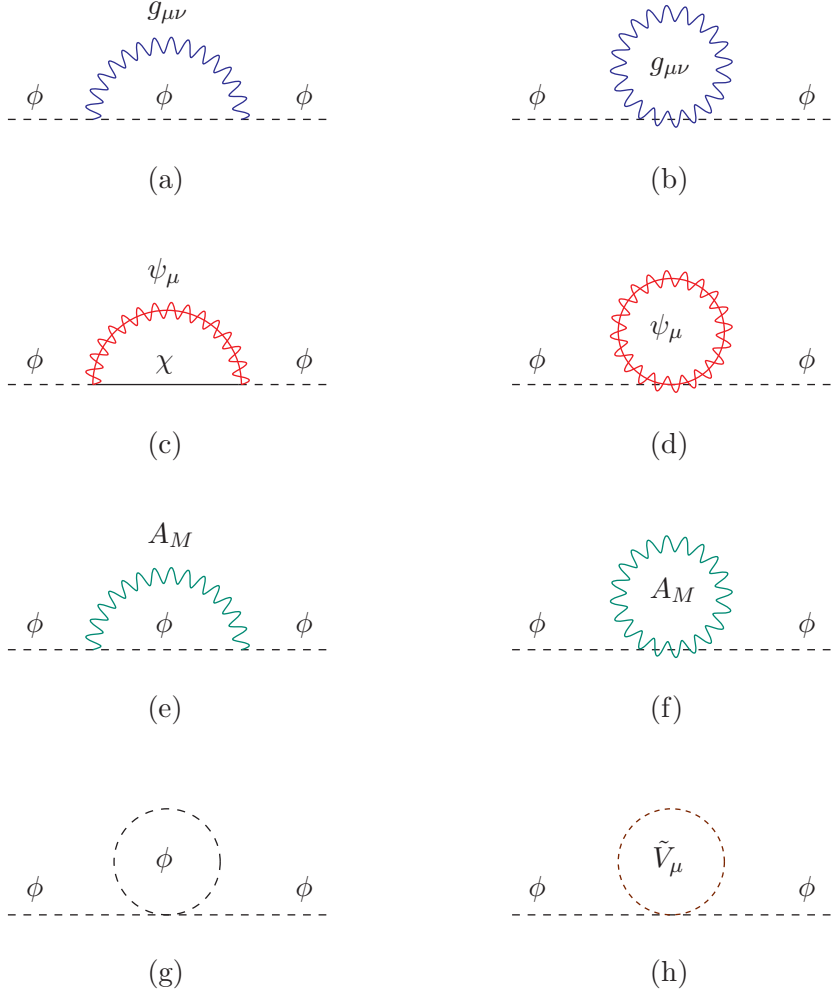
\begin{figure}[h]
\begin{center} 
\begin{picture}(355,390)(15,-25)
\put(42,307){$\phi$}
\put(144,307){$\phi$}
\put(91,307){$\phi$}
\DashLine(35,300)(155,300){3}
\SetColor{Blue}
\PhotonArc(95,300)(28,0,180){3}{15}
\put(88,341){$g_{\mu\nu}$}
\SetColor{Black}
\put(88,275){(a)}
\put(232,307){$\phi$}
\put(334,307){$\phi$}
\DashLine(225,300)(345,300){3}
\SetColor{Blue}
\put(278,320){$g_{\mu\nu}$}
\PhotonArc(285,320)(20,0,360){3}{20}
\SetColor{Black}
\put(278,275){(b)}
\put(42,207){$\phi$}
\put(144,207){$\phi$}
\put(91,207){$\chi$}
\put(88,241){$\psi_{\mu}$}
\DashLine(35,200)(68,200){3}
\Line(68,200)(122,200)
\DashLine(122,200)(155,200){3}
\SetColor{Red}
\CArc(95,200)(28,0,180)
\PhotonArc(95,200)(28,0,180){3}{15}
\SetColor{Black}
\put(88,175){(c)}
\put(232,207){$\phi$}
\put(334,207){$\phi$}
\put(278,220){$\psi_{\mu}$}
\DashLine(225,200)(345,200){3}
\SetColor{Red}
\CArc(285,220)(20,0,360)
\PhotonArc(285,220)(20,0,360){3}{20}
\SetColor{Black}
\put(278,175){(d)}
\put(42,107){$\phi$}
\put(144,107){$\phi$}
\put(91,107){$\phi$}
\put(88,141){$A_{M}$}
\DashLine(35,100)(155,100){3}
\SetColor{PineGreen}
\PhotonArc(95,100)(28,0,180){3}{15}
\SetColor{Black}
\put(88,75){(e)}
\put(232,107){$\phi$}
\put(334,107){$\phi$}
\put(278,120){$A_{M}$}
\DashLine(225,100)(345,100){3}
\SetColor{PineGreen}
\PhotonArc(285,120)(20,0,360){3}{20}
\SetColor{Black}
\put(278,75){(f)}
\put(42,7){$\phi$}
\put(144,7){$\phi$}
\put(91,18){$\phi$}
\DashLine(35,0)(155,0){3}
\DashCArc(95,20)(20,0,360){3}
\put(88,-25){(g)}
\put(232,7){$\phi$}
\put(334,7){$\phi$}
\put(279,18){$\tilde V_{\mu}$}
\DashLine(225,0)(345,0){3}
\SetColor{Brown}
\DashCArc(285,20)(20,0,360){2}
\SetColor{Black}
\put(278,-25){(h)}
\end{picture}
\caption{\em The diagrams contributing to the mass of the scalar $\phi$.}
\label{diagcanc}
\end{center}
\end{figure}

The 8 diagrams contributing to the one-loop mass correction are depicted in Fig.~\ref{diagcanc}. 
As in the rigid case~\cite{Mirabelli}, the singular couplings proportional to $\delta(0)$ 
play a crucial r\^ole in the supersymmetric cancellation. Notice however that 
the auxiliary field $\tilde V_\mu$ gives a non-vanishing contribution as well, which is in fact 
the only contribution left over in the effective action when integrating it out. 
Using the representation~\cite{Mirabelli}
\bea
e_{\dot 5}^5\,\delta(0) = \frac 1{2\pi R} \sum_{n=-\infty}^\infty \frac {p^2 - m_n^2}{p^2 - m_n^2} \;.
\eea
all the diagrams can be brought into the form
\be
\Delta m^2_\alpha = \frac i{2 \pi R} \sum_{n=-\infty}^\infty \int \frac {d^4p}{(2 \pi)^4} 
\frac {{N_\alpha}}{p^2 - m_n^2} \;.
\ee
After a straightforward computation, one can verify that the diagrams indeed cancel 
each other level by level, the contributions of the single diagrams being\footnote{We
believe that this corrects the computation performed in ref.~\cite{Gherghetta}, where 
the diagrams (f) and (h) where not properly taken into account, as well as that of
\cite{Antoniadis}.}:
\bea
\begin{array}{lcllcl}
N_a \a=\a 0 \;,\a
N_b \a=\a \displaystyle{\frac 53}\, p^2 \;, \smallskip \\
N_c \a=\a 0 \;,\a
N_d \a=\a - \displaystyle{\frac 83}\, p^2 \;, \smallskip \\
N_e \a=\a \displaystyle{\frac 13}\,(p^2 - m_n^2) \;,\a
N_f \a=\a - \displaystyle{\frac 13}\,(p^2 - 4\, m_n^2)\;, \smallskip \\
N_g \a=\a - \displaystyle{\frac 13}\,(p^2 - m_n^2) \;,\;\;\a
N_h \a=\a \displaystyle{\frac 43}\,(p^2 - m_n^2)\;.
\end{array}
\label{canc}
\eea
The diagrams $(a)$ and $(c)$ involving cubic vertices vanish, since the graviton or 
gravitino going out of a cubic vertex turns out to be longitudinal so that it cannot couple 
to an other cubic vertex. Indeed, one can easily verify that
$(p^2 \eta^{\mu\nu} - p^\mu p^\nu) \langle h_{\mu\nu} h_{\alpha\beta}\rangle \propto p_\alpha p_\beta$
and $\gamma^{\mu\nu} p_\mu \langle \Psi_\nu \bar \Psi_\alpha\rangle \propto p_\alpha$.
The singular diagram $(g)$ arising from the quartic scalar coupling proportional to $\delta(0)$ 
cancels the divergent part of diagrams $(e)$, $(f)$ and $(h)$, similarly to what 
happens in the rigid case~\cite{Mirabelli}. Actually, the diagrams in the left column ($a$, $c$ $(e)$ 
and $(g)$) which involve virtual matter particles cancel separately.
This is because the theory with frozen matter fields, where there are only the diagrams on the right column,
is a consistent construction on its own (see section \ref{stab}), for which the 
cancellation must hold true as well. 

It is expected that this pattern of cancellation will continue for operators with higher 
powers of scalar fields. Unlike what happens in the rigid case~\cite{Mirabelli},
expanding the Lagrangian~(\ref{L}) to higher powers in $\phi$ generates higher powers 
of $\delta(0)$. The associated singular scalar diagrams are expected to contribute to 
cancel the divergences coming from the graviphoton, but we will not proceed further.

\subsection{Partially off-shell formulation}

In the partially off-shell formulation defined by (\ref{Ltilde}), things are easier, and one 
can verify the supersymmetric cancellation of the full effective scalar potential. 
The graviton and gravitino propagators are exactly as before. In this case, the graviphoton 
does not couple to matter, and correspondingly singular self-couplings for matter fields are absent.
The propagator of the auxiliary vector field $V_\mu$ is in this case non-trivial, as a consequence 
of its mixing with the graviphoton, and inverting (\ref{KVA}) one easily finds:
\bea
\langle V_\mu V_\nu \rangle_n \a=\a -3\,\Big[p^2\eta_{\mu\nu} - p_\mu p_\nu\Big] \frac i{p^2 - m_n^2} \;.
\eea

As before, cubic vertices involving gravitons and gravitinos are irrelevant, and cubic vertices 
involving the vector field vanish trivially at zero-momentum due to the fact that its propagator 
is transverse. The relevant diagrams are then loops of gravitons, gravitinos or vector fields, with 
an arbitrary number of insertions of the appropriate quartic vertex with scalar fields. 
In order to perform an exact resummation of all these one-loop diagrams, it is extremely convenient 
to introduce the following projection operators:
\bea
P_{1/2}^{\mu\nu} \a=\a \frac 13 \Big(\gamma^\mu \!- \frac {p^\mu}{p\!\!\!/\,}\Big)
\Big(\gamma^\nu \!- \frac {p^\nu}{p\!\!\!/\,}\Big) \;, \\
P_{1}^{\mu \nu} \a=\a \eta^{\mu\nu} \!- \frac {p^\mu p^\nu}{p^2} \;, \\
P_{3/2}^{\mu \nu} \a=\a \Big(\eta^{\mu\nu} \!- \frac {p^\mu p^\nu}{p^2} \Big)
- \frac 13 \Big(\gamma^\mu \!- \frac {p^\mu}{p\!\!\!/\,}\Big)
\Big(\gamma^\nu \!- \frac {p^\nu}{p\!\!\!/\,}\Big) \;, \\
P_{2}^{\mu \nu \alpha \beta} \!\! \a=\a 
\frac 12 \Big(\eta^{\mu\alpha} \!- \frac {p^\mu p^\alpha\!}{p^2} \Big)
\Big(\eta^{\nu\beta} \!- \frac {p^\nu p^\beta\!}{p^2} \Big) 
- \frac 16 \Big(\eta^{\mu\nu} \!- \frac {p^\mu p^\nu\!}{p^2} \Big)
\Big(\eta^{\alpha\beta} \!- \frac {p^\alpha p^\beta\!}{p^2} \Big) + (\alpha \leftrightarrow \beta) 
\;.\hspace{20pt}
\eea
These are all idempotent, $P_{i}^2 = P_{i}$, and transverse, $p \cdot P_{i}=0$. The 
spin-$3/2$ projector also satisfies $\gamma \cdot P_{3/2} = 0$. Defining for notational
convenience $\rho = \frac 13 |\phi|^2$, the quartic interaction vertices in mixed momentum/configuration 
space can then be written as 
\bea
\Lag^{\rm int} \a=\a \rho\, \delta(x^5) 
\Big[p^2 h_{\mu\nu} \Big(P_{2}^{\mu \nu \alpha \beta}\!
- \frac 23 P_{1}^{\mu \nu} P_{1}^{\alpha\beta}\Big) h_{\alpha\beta} \nn \\
\a\;\a \hspace{35pt} +\, 2\, \bar \Psi_\mu\, p\!\!\!/\,\Big(P_{3/2}^{\mu \nu} 
- 2 P_{1/2}^{\mu\nu}\Big) P_L \Psi_\nu + \frac 13 V_\mu V^\mu \Big] \;.
\label{Vquart}
\eea
The longitudinal parts of the graviton and gravitino propagators are irrelevant. 
It is then convenient to use this fact and choose the longitudinal part in such a 
way as to reconstruct for each propagator the appropriate projection operator, 
respectively $P_{2}^{\mu \nu \alpha \beta}$ and $P_{3/2}^{\mu \nu}$. 
The vector propagator,
happily, is already proportional to the projection operator $P_{1}^{\mu\nu}$. 
Furthermore, the mass insertion in the gravitino propagator drops in the diagrams because of the 
$P_L$ projectors at the vertices. 
In practice, one can therefore use the following propagators:
\be
\Delta_{(h)}^{\mu\nu \alpha\beta} = P_{2}^{\mu\nu\alpha\beta}\,\Delta \;,\qquad
\Delta_{(\Psi)}^{\mu\nu} = \frac 12\, p\!\!\!/\, P_{3/2}^{\mu\nu}\, \Delta \;, \qquad
\Delta_{(V)}^{\mu\nu} = 3\,p^2 P_{1}^{\mu\nu}\, \Delta \;.
\ee
where 
\be
\Delta = \frac 1{2\pi R} \sum_{n=-\infty}^\infty \frac i{p^2 - m_n^2} \;.
\ee
Since $P_{1} \perp P_{2}$ and $P_{1/2} \perp P_{3/2}$, the quartic vertex acting on the graviton and 
gravitino propagators is just proportional to respectively $P_2$ and $P_{3/2}$. The effective potential 
is then easily computed by resumming insertions in the graviton, gravitino and graviphoton 
vacuum diagrams. One finds
\bea
W_{h+\psi+A}(\rho) \a=\a -\frac 12 \int \! \frac{d^4p}{(2\pi)^4} 
\sum_{k=1}^\infty \frac {(-i\rho\,p^2)^k}k {\rm Tr}\, \Big[(P_{2}\,\Delta \big)^k 
- 2 (P_{3/2}P_R\, \Delta \big)^k + (P_{1}\, \Delta \big)^k \Big] \nn \\
\a=\a \Big({\rm Tr}\,P_{2} - {\rm Tr}\,P_{3/2} + {\rm Tr}\,P_{1} \Big)\,
\frac 12 \int \! \frac{d^4p}{(2\pi)^4} {\rm ln}\,\Big[1 + i \rho\, p^2 \Delta\Big] \;.
\eea
The vanishing of the one-loop effective potential is thus a direct consequence of the standard 
balancing of degrees of freedom in supergravity: ${\rm Tr}\,P_{2} - {\rm Tr}\,P_{3/2} 
+ {\rm Tr}\,P_{1} = 5 - 8 + 3 =0$. The quantity multiplied by this coefficient is easily 
recognized to be the effective potential induced by a real scalar field $\varphi$, corresponding 
to a single degree of freedom, with the following Lagrangian:
\be
\Lag_\varphi = \partial_M \varphi \partial^M \varphi 
+ \rho\, \delta(x^5)\partial_\mu \varphi \partial^\mu \varphi \;.
\ee
Indeed, defining $f_n = i/(p^2 - m_n^2)$, in terms of which $\Delta = (2 \pi R)^{-1} \sum_n f_n$,
one computes
\bea
W_{\varphi}(\rho) \a=\a \frac 12\, {\rm ln}\,{\rm det} 
\Big[1 -  \rho\,\delta(x^5)\,\frac {\partial_\mu \partial^\mu}{\partial_M \partial^M}\Big] =
\frac 12 \int \! \frac{d^4p}{(2\pi)^4} {\rm ln}\,{\rm det}_{\rm KK} 
\Big[\delta_{n,n^\prime} + \frac {i \rho\, p^2}{2 \pi R} f_n\Big] \nn \\
\a=\a \frac 12 \int \! \frac{d^4p}{(2\pi)^4} {\rm ln}\, 
\Big[1 + \frac {i \rho\, p^2}{2 \pi R}\,\sum_n f_n\Big] 
= \frac 12 \int \! \frac{d^4p}{(2\pi)^4} {\rm ln}\, 
\Big[1 + i \rho\, p^2\,\Delta \Big] \;.
\eea
The determinant over the infinite KK modes (needed in the third equality) is most easily computed 
by considering recursively finite truncations of increasing dimensionality.

\setcounter{equation}{0}
\section{Scherk--Schwarz supersymmetry breaking}
\label{SS}

We want to consider a situation where supersymmetry is broken by the VEV of the radion 
auxiliary field. As argued in~\cite{MartiKaplan}, this case corresponds to Scherk--Schwarz 
supersymmetry breaking. This correspondence has been further elucidated in~\cite{vonGersdorff1} 
by considering the off-shell formulation of 5D supergravity. Furthermore the same supersymmetry 
breaking spectrum has been obtained in~\cite{Bagger} by considering constant superpotentials 
localized at the fixed-points. The latter realization can be simply understood in the effective 
field theory. The boundary term  leads to a constant 4D superpotential so that eq.~(\ref{effoff})
becomes
\be
\Lag^{\rm eff}=\Big[(T+T^\dagger)\Big({\sqrt{S_0\,S_0^\dagger}}+\frac 12 \Big)\Big]_D +
P\Bigl [S_0\Bigr ]_{\cal F} + P^*\Bigl [S_0\Bigr ]_{\cal F}^\dagger\;,
\label{effconst}
\ee
corresponding to the following structure as far as the auxiliary fields are concerned
\be
\Lag^{\rm eff}_{F_{S_0,T}} = (T+T^*)|F_{S_0}|^2
+(F_T+P^*) F_{S_0}^*+(F_T^*+P) F_{S_0} \;.
\label{effconst2}
\ee
Solving the auxiliary equations of motion we find the standard no-scale result: 
$F_{S_0}=0$, $F_T = - P^*$, with the scalar potential exactly zero for any $T$.

For the purpose of our calculation, as it will become clear below, it is important
to understand in some detail the way $F_T$ is generated in the full 5D theory. 
From the discussion in section 3 we have
\be
F_T=\frac{1}{2}\int_{-\pi}^{\pi} d x^5 \left [E_5^{\dot 5}\right ]_F=
\frac{1}{2}\int_{-\pi}^{\pi} d x^5\left [V^1_5+iV^2_5 +4e_5^{\dot 5}(it^1-t^2)\right] \;.
\ee 
Notice that all components of $E_5^{\dot 5}$ can be locally gauged away, so that when 
$F_T\not =0$ supersymmetry is broken by global effects at the compactification scale. 
This is very similar to what happens for a $\hbox{U}(1)$ gauge symmetry in the presence of localized 
Fayet-Iliopoulos terms~\cite{Mirabelli}.
We are interested in the situation in which $F_T$ is the only auxiliary with non zero VEV.
Therefore, $t^1$ and $t^2$, which are part of the gravitational multiplet should vanish
and we have just $F_T\propto V^1_5+iV^2_5$. To generate $F_T$ we add boundary superpotentials
\cite{Bagger} in our off-shell formulation. The superpotential being a complex object, 
there are two independent real covariant densities that we can write at each 
boundary~\cite{Sohnius,Zucker}:
\bea
{\rm Re}\left[S_0\right]_{F} \a=\a \frac{1}{2} \bar\Psi_a \gamma^{ab}(Y^1\tau_1 + Y^2\tau_2)\Psi_b
-2N - 12 (Y^2 t^2+Y^1t^1) + D_{\dot 5} Y^3 +\dots  \;, \label{superboundary} \\
{\rm Im}\left[S_0\right]_{F} \a=\a \frac{1}{2}\bar \Psi_a\gamma^{ab}(Y^1 \tau_2 - Y^2 \tau_1) \Psi_b 
+ 2 W^{\dot 5} + \dots \;. \label{Gboundary}
\eea
In both equations the dots indicate $\rho$-dependent terms, which trivially vanish on-shell 
and can thus be discarded. ${\rm Re}\,F_{S_0}$ and ${\rm Im}\,F_{S_0}$ are fairly different objects 
when written in terms of 5D fields, see eqs.~(\ref{FS0}) and (\ref{GS0}). Because of that, 
there are important technical differences in working out the implications of adding the 
${\rm Re}\,F$ and the ${\rm Im}\,F$ terms. In the next two subsections we will separately 
study the two cases.

\subsection{Generating $V_5^2$} 
\label{generateV2}

Let us consider adding to the action a superpotential term
\be
{\cal L}_\epsilon = -P_\epsilon(x^5){\rm Im}[S_0]_F
\label{S0G} 
\ee 
where
\be
P_\epsilon(x^5) = 2\pi\epsilon_0\,\delta(x^5) + 2\pi\epsilon_\pi\,\delta(x^5-\pi) \;.
\ee
Using eq.~(\ref{WM}) and writing eq.~(\ref{Gboundary}) in terms of $B_{MNR}$ and $\Psi_M$ 
we immediately encounter a problem. The gravitino bilinear  cancels out and what remains 
is just a total derivative:
\be
{\rm Im}[S_0]_F= \frac{1}{6}\epsilon^{\dot 5\mu\nu\rho\sigma}\partial_\mu B_{\nu\rho\sigma} \;.
\ee
Naively this term is trivial, though a more correct statement is that it is topological, as it 
can be formally associated to an integral at the boundary of our 4D space (not the boundaries 
of the orbifold!). This result indicates that, as it stands, the off-shell Lagrangian 
with a tensor multiplet compensator of ref. \cite{Zucker} is not fully adequate to describe this 
particular superpotential. In deriving the Lagrangian no attention was paid to total derivative 
terms. Now, the fact that for certain auxiliary formulations of supergravity, some ways of 
breaking supersymmetry are triggered by global, instead of local, charges is known\footnote{We 
thank C. Kounnas for pointing this out to us.}. The basic point is that the set of auxiliary 
fields we are using is perfectly fine locally, but there can be physical situations where a 
global definition of our fields, in particular $B_{MNR}$, is impossible and our set of fields 
inadequate. This is the analogue of what happens for monopole configurations of a gauge vector 
field. These are the situations where there is a non-zero 4D-flux for $dB$. This may not be a 
big surprise. The tensor $B$ was originally introduced to locally solve the constraint 
on the vector of a linear multiplet. After gauge-fixing, this constraint reads:
\be
\partial_M W^M + \partial_M J_\Psi^{M} = 0 \;,
\label{constr}
\ee
in terms of the $\hbox{U}(1)_{T_2}$ gravitino current
\be
J_\Psi^{M} = - \frac 14 \bar \Psi_A \gamma^{AMB} \tau_2 \Psi_B \;.
\ee
Using the language of differential forms, eq.~(\ref{constr}) reads $d^{*} (W + J_\Psi) = 0$, and 
this is solved by eq.~(\ref{WM}) with $\rho = 0$: $W = - J_\Psi + \frac 1{12} {}^*dB$. When the 
space has non trivial 4-cycles this parametrization is missing the closed 4-forms $\omega$ which 
are not exact $\omega \not = d B$, but which are perfectly acceptable solutions of the constraint.
Fortunately, for the purpose of our computations, this lack of completeness is not posing any 
serious limitations. This will become clear in the following discussion. It would nevertheless
be very interesting to address this issue within the potentially more powerful formalism developed
in~\cite{Kugo,Bergshoeff}.

From inspection of the low energy effective theory, eq.~(\ref{effconst2}), the superpotential 
${\rm Im}\,F$-term we are considering would correspond to imaginary $P$ and would induce a VEV for 
${\rm Im} F_T\propto V_5^2$. The terms in the bulk Lagrangian (eq.~(\ref{L5ini})) that are 
relevant to discuss the VEV of $V_5^2$ and its consequences are
\bea
{\cal L} = - \frac i2 \bar\Psi_M \gamma^{MNP}{D}_N \Psi_P + V_M^2 J_\Psi^M
- \frac 1{12} \epsilon^{ABMNP}\partial_A V_B^2 B_{MNP} +W_A W^A.
\label{Lini}
\eea
Integrating by parts and using the definition (\ref{WM}) of $W^M$ (with $\rho = 0$), this 
equation becomes 
\bea
{\cal L} = - \frac i2 \bar\Psi_M \gamma^{MNP} D_N \Psi_P + W^A V_A^2+W^A W_A \, .
\label{Lmix}
\eea
Notice that the coupling between $V_M^2$ and the gravitino no longer shows up explicitly.
Eq.~(\ref{Lmix}) has precisely the structure of the no-scale Lagrangian (\ref{effconst2}), 
in particular there is no $V_M^2V^{M2}$ term. If we treat $W_M$ as an independent field, 
the sum of eqs.~(\ref{Lmix}) and (\ref{S0G}), integrated over $x^5$ and reduced to the zero 
modes ${\rm Im}\,F_T=\pi V_5^2$, ${\rm Im}\,F_{S_0}=2W^{\dot 5}$, agrees perfectly with eq.~(\ref{effconst2}).
However, from the 5D point of view $W^M$ cannot be independent, otherwise the gauge symmetry 
$\hbox{U}(1)_{T_2}$ would be explicitly broken; the constraint (\ref{constr}) represents precisely
the condition for (\ref{Lmix}) being gauge-invariant. These considerations will matter in moment. 
Before then, let us consider the equations of motion that follow form the unconstrained fields 
in eq.~(\ref{Lini}). The equation for $V_A^2$ gives $W^A=0$, so that the equation for $B_{MNR}$ 
implies $\partial_M V_N^2-\partial_N V_M^2=0$. Up to gauge transformations, the resulting class 
of solutions is conveniently parametrized by a constant $V_5^2=2\epsilon$~\cite{Zucker,vonGersdorff1}. 
Indeed the corresponding gauge-invariant Wilson line operator is $e^{i\!\oint dx^5 V_5^2 \tau_2/2}$ 
so that physics should be unchanged by the shift $\epsilon\to \epsilon+1$. Going back to the 
explicit Lagrangian eq.~(\ref{Lini}), we find a gravitino mass $\propto \epsilon$, in the correct 
relation with $F_T$, see eq.~(\ref{noscalemass})\footnote{Actually there is a subtlety in 
deriving this mass term, which is related to the fact that our formulation is not completely 
satisfactory at the global level. The equation of motion of $V_2$ sets $W = 0$, i.e. 
${}^*dB = 12 J_\Psi$. This does not just fix the value of the auxiliary field $B$; when integrated 
over a 4D surface, it also gives the condition $\int_4 J_\Psi^5 = 0$. Since this is a constraint 
on the physical fields, the Lagrangian obtained by substituting the solution for the auxiliary fields 
would miss the terms associated to the constraint: substituting $W=0$ into eq.~(\ref{Lmix}) one
finds no gravitino mass term at all. The correct procedure is to first derive the equation of 
motion of the gravitino, and then solve the constraint from the auxiliary fields.}. 
The existence of this family of 
supersymmetry breaking solutions is in direct  correspondence with eq.~(\ref{S0G}) being a total 
derivative. The compensator auxiliary field $W^{\dot 5}$ is not the most general scalar: it is 
basically the field strength of a 3-form (modulo the gravitino term). The variation of $W^{\dot 5}$ 
imposes a slightly weaker constraint than usual. The actual value of $\epsilon$ cannot be decided 
with the sole use of our local description: the source of $\epsilon$ is a global flux. However 
for the purpose of our computation all we need is a {\it locally} consistent way to generate 
$F_T\not =0$. We have just shown that the {\it local} Lagrangian \cite{Zucker} we use admits 
automatically these solutions, though it formally lacks the {\it global} degrees of freedom needed
to associate $\epsilon$ to a Lagrangian parameter (a charge)\footnote{In ref.~\cite{vonGersdorff2} 
the point of view was taken taken that the 1-loop effective action should be minimized with respect 
to $\epsilon$. However, as $V_M^2$ satisfies the equation $\partial_N V_M^2-\partial_M V_N^2=0$,
there is no local propagating degree of freedom associated to the VEV $V_5^2=2\epsilon$. There is 
no dynamics that can make $\epsilon$ evolve locally, and  we do not fully understand the meaning 
of that minimization. Our viewpoint is that in the correct treatment $V_5^2$ should be fixed at 
tree level by a global charge, so that the issue of minimizing in $\epsilon$ should  not arise.}.

The bottom line of the above discussion is that in terms of $W^M$ the 5D Lagrangian looks precisely 
like what one would have liked, and reproduces nicely the 4D structure. But in terms of $B_{MNR}$ 
there are differences. In fact if one could do without $B_{MNR}$ and just work with a constrained 
$W^M$ these issues would not arise: the most general ${}^*W$ includes closed forms with non-zero flux. 
Unfortunately the fully off-shell Lagrangian (\ref{simplified}) cannot be written just in terms of 
$W^M$, not even after integrating by parts. Indeed all the obstruction is coming from the second 
to last term in eq.~(\ref{simplified}). This problem is fully analogous to the case of ${\cal N}=2$ 
supergravity in 4D which was discussed in ref.~\cite{deWit}. The basic remark is that the Lagrangian 
can be written in terms of $W$, but at the price of loosing manifest $\hbox{SU}(2)_R$ invariance. Now, 
after gauge fixing $Y_i\propto \delta_i^2$ the obstructive term in eq.~(\ref{simplified}) 
vanishes, and we can write the Lagrangian just in terms of $W^M$, i.e. eq.~(\ref{Lmix}). Like in 
ref.~\cite{deWit} we can enforce the constraint on $W^M$ by adding a Lagrange multiplier $X$
\be
{\cal L}_X = \partial_M X\left(W^M+J_\Psi^M\right) \;.
\label{multiplier}
\ee
Notice that $X$ shifts under $\hbox{U}(1)_{T_2}$ gauge rotations, restoring invariance of the unconstrained 
Lagrangian (\ref{Lmix}). Now, the addition of eqs.~(\ref{S0G}), (\ref{Lmix}) and (\ref{multiplier}) 
leads to the equations of motion $W^M = 0$ and $V_M^2 = 2 P_\epsilon(x^5)\delta_M^5+\partial_M X$.
The latter equation is manifestly gauge invariant, and fixes just the VEV of the Wilson line. Defining
$\epsilon = \epsilon_1 + \epsilon_2$ we find:
\be
\oint dx^5 V_5^2 = 2 \oint dx^5 P_\epsilon(x^5) = 4\pi \epsilon \;.
\ee
Notice also that on shell the gravitino mass is still determined by $V_5^2$. So a convenient gauge to 
study the gravitino spectrum is the one in which $V_5^2=2\epsilon$ is constant, and there are no 
$\delta$-function terms. We will better explain below the advantages of working in a gauge with 
no $\delta$-function terms. We see that in this approach with $W$ instead of $B$ we end up with 
the same conclusions.

Let us now study the spectrum and the wave-function of the gravitino in the presence of the 
Wilson line. With a constant $V_5^2 \neq 0$, the zero mode of $\psi_5^2$ plays the role of the Goldstino, 
so it can be gauged away. The gravitini can be described through a doublet of Weyl spinors 
$\chi_\mu = (\chi_\mu^1,\chi_\mu^2)^T$. They can then be decomposed as 
$\chi^\mu(x^5) = \sum_n \xi_n (x^5) \chi_n^\mu$ in terms of Weyl KK modes $\chi_n^\mu$ and 
the standard wave-functions
\be
\xi_n(x^5) = \left(\matrix{\cos n x^5  \cr 
-\sin n x^5}\right) \;.
\label{wave}
\ee
The mass eigenstates are Majorana KK modes defined as $\psi_\mu^n = (\chi_\mu^n, \bar \chi_\mu^n)^T$,
with masses given by
\be
m_n(\epsilon) = \frac {n-\epsilon}R \;.
\ee
Notice that the periodicity $\epsilon\to \epsilon+1$ is respected. Moreover, for the $n=0$ mode, 
which for $\epsilon \ll 1$ represents the 4D gravitino, we reproduce the well known relation of 
the no-scale model
\be
m_{3/2}=\Big|\frac{F_T}{T+T^*}\Big|=\Big|\frac{\pi V_5^2}{2\pi R}\Big|=\Big|\frac{\epsilon}{R}\Big| \;.
\label{noscalemass}
\ee
Notice also that the wave functions of the modes are unaffected by supersymmetry breaking. 
In particular they are smooth at the boundaries. It will become clear below why this matters.

The above scenario is shown to be equivalent to Scherk--Schwarz supersymmetry breaking by performing a
non-single valued $\hbox{U}(1)_{T_2}$ gauge transformation $e^{-2 i \alpha_2 T_2}$ to eliminate $V_5^2$,
in the spirit of ref.~\cite{Hosotani}. This is achieved with $\alpha_2(x^5) = \epsilon x^5$. 
In this new basis, $V_5^{\prime 2} = V_5^2 - 2\epsilon = 0$, but the new charged fields 
$\phi^\prime = e^{- 2 i \epsilon x^5 T_2} \phi$ get twisted boundary conditions.
Defining the matrices $U(\epsilon) = e^{- 4 i \epsilon T_2}$ and $Z = \eta_\phi (-1)^{T-T_3}$ 
for any given  multiplet $\phi$ with isospin $T$ and overall parity $\eta_\phi$, the new 
boundary conditions are
\bea
\phi^\prime(x^5 + 2\pi ) =  U(\epsilon)\phi^\prime(x^5) \;,\;\;
\phi^\prime(-x^5)= Z \phi^ \prime(x^5) \;.
\eea
The ${\rm SU}(2)_R$ group algebra ensures that the consistency condition 
$Z U(\epsilon) Z= U(\epsilon)^{-1}$ is automatically satisfied.

In the primed basis it is manifest that supersymmetry is broken non-locally. It amounts to 
the fact that the two different fixed-point locally preserve different combinations of the 
supercharges. Indeed, the reflection condition of the new gravitino around each fixed point 
$x^5=k\pi$ involves a different matrix $Z(\epsilon,k) = e^{-ik\pi \epsilon \tau_2} \tau_3 
e^{ik\pi \epsilon \tau_2}$ and reads
\be
\chi_\mu^\prime(k\pi+y) = Z(\epsilon,k) \chi^\prime_\mu(k\pi-y)\, .
\ee
The supersymmetry that is locally preserved at $x^5=k \pi$ is aligned with the $+1$ eigenvalue 
of $Z(\epsilon,k)$. The fields diagonalizing the latter are nothing but the gravitini in the 
unprimed basis, $e^{i k\pi \epsilon \tau_2} \chi_\mu^\prime(k\pi) = \chi_\mu(k\pi)$, and the 
combination of gravitini associated with the supersymmetry preserved at $x^5=k\pi$ is thus
$\cos k\pi\epsilon\, \chi^{\prime 1}_\mu(k\pi)
+\sin k\pi\epsilon\, \chi^{\prime 2}_\mu(k\pi) = \chi_\mu^1(k\pi)$.
Therefore, working in the Scherk--Schwarz picture, i.e. in the primed basis, one has to be 
careful when writing boundary actions to use the right combination of the two gravitini. 
The couplings are straightforward in the unprimed basis where the fields are single valued. 
In this basis the wave-functions of the appropriate gravitino components at each fixed point 
are therefore given by:
\be
\chi^\mu (0) = \sum_{n=-\infty}^\infty 
\left(\matrix{1 \cr 0} \right) \chi^\mu_n \;,\;\;
\chi^\mu (\pi) = \sum_{n=-\infty}^\infty 
\left(\matrix{1 \cr 0} \right) (-1)^n \chi^\mu_n \;.
\label{wavefactors}
\ee
Notice that no dependence on $\epsilon$ appears in these couplings. In ref.~\cite{Gherghetta} 
this dependence was not eliminated, leading, in general, to  incorrect results. We will have 
more to say on this issue in the next section.

Summarizing: {\em the net effect of a Scherk--Schwarz twist in the 
five-dimensional theory amounts to a shift  in the masses of the gravitino KK modes}. 
The gravitino wave-functions that determine the couplings to the fixed-points
are instead insensitive to the twist and coincide with those of the supersymmetric case; the 
$n$-th mode has therefore wave-function $1$ at $x^5=0$ and $(-1)^n$ at $x^5 = \pi$. 

\subsection{Generating $V_5^1$}

Consider now the addition of a superpotential ${\cal L}_\eta = -P_\eta(x^5){\rm Re}[S_0]_F$, with 
${\rm Re}[S_0]_F$ given by eq.~(\ref{superboundary}) and 
\be
P_\eta(x^5) = 2\pi\eta_0\,\delta(x^5) + 2\pi\eta_\pi\,\delta(x^5-\pi) \;.
\ee
In the gauge $Y^{1,3}=0$, $Y^2=1$, eq.~(\ref{superboundary}) reproduces the structure of eq.~(\ref{effconst}) 
for real $P$. The auxiliary field $N$ is an ordinary scalar, so that the problem of the previous
section does not arise. Let us consider then the equations of motion in the presence of this 
superpotential. As we have already discussed in section~\ref{partoff} by introducing 
$N_\pm = N + 6 t^2 \pm \frac 12 V_{\dot 5}^1$, the auxiliary fields do not contribute any 
term to the on-shell action, because their equation of motion implies $N_-=0$. 
In terms of the original fields $t^2$, $V_5^1$ and $N$, one finds $t^2=0$, $V_5^1 =  2 P_\eta(x^5)$ 
and $N = P_\eta(x^5)/R$. In particular, a non-vanishing VEV for the zero mode of $V_5^1$ is 
generated; defining $\eta = \eta_0 + \eta_\pi$, we have
\be
\oint dx^5 V_5^1 = 2\oint dx^5 P_\eta(x^5) = 4\pi \eta \;.
\ee

Let us now study the gravitino spectrum and wave functions.
The mass operator for the doublet of Weyl spinors $\chi_\mu = (\chi_\mu^1,\chi_\mu^2)^T$ 
describing the gravitino is given by the matrix
\be
M = \left(\matrix{-i P_\eta(x^5) \a -\partial_5 \cr \partial_5 \a 0\cr}\right) \;.
\label{massoperator}
\ee
This leads to singularly behaved wave functions at the boundary~\cite{Bagger}: $\chi_\mu^1$ has 
a cusp and $\chi_\mu^2$ is discontinuous. As it was the case in the Scherk--Schwarz example, this 
situation leads to ambiguities when trying to decide which combination couples to matter at the 
boundary. Notice however that this singular behavior comes along with a singular profile 
$V_5^1(x^5) = 2 P_\eta(x^5)$. Therefore, a natural guess is that by going to a gauge in which 
$V_5^1$ is smooth, the gravitini will also be smooth, and their interactions straightforward. 
This is indeed what happens. What we need is a $\hbox{U}(1)_{T_1}$ rotation $e^{-2 i \alpha_1 T_1}$ 
with a parameter $\alpha_1$ such that $V_5^1$ is made constant. One finds
\be
\alpha_1(x^5)=-\epsilon(x^5)\left [\eta_0(|x^5|-\pi)+\eta_\pi |x^5|\right ] \;,
\label{goodgauge}
\ee
where $\epsilon(x^5)$ is the completely odd step function which jumps from $-1$ to $1$ at 
$x^5=2k\pi$ and from $1$ to $-1$ at $x=(2k+1)\pi$. Notice that $\alpha_1$ is defined to be 
single valued on the circle, although it is discontinuous at the fixed points. This should 
be contrasted to the improper gauge transformation of the previous section. In the new gauge, 
the gravitino is transformed to a new field $\Psi_\mu^\prime$, the gauge field is shifted to
\be
V_5^{\prime 1}=V_5^1 - 2 \partial_5 \alpha_1 = 2\eta
\label{V1smooth}
\ee
and $Y_2$ and $Y_3$ are rotated to
\bea
Y^{\prime 2}\a=\a\cos 2\alpha_1 Y^2-\sin 2\alpha_1Y^3= \cos 2\alpha_1 \;, \\
Y^{\prime 3}\a=\a\cos 2\alpha_1 Y^3+\sin 2\alpha_1Y^2= \sin 2\alpha_1 \;.
\label{rotate}
\eea
In order to compute the gravitino mass term in this new gauge, one has to go back to 
eq.~(\ref{simplified}), where invariance under the full ${\rm SU}(2)_R$ is still manifest. 
In particular, one has to consider the last term in eq.~(\ref{simplified}), which, due to 
the non-constant profile for $Y^i$, gives rise to an extra contribution to the gravitino 
mass operator\footnote{Notice that at the same time the covariant derivative 
${\cal D}_M^\prime$ in eq.~(\ref{L5ini}) will involve the rotated vector 
$V_M^{\prime 2} = \cos 2 \alpha_1 V_M^2 - \sin 2 \alpha_1 V_M^3$. This does not influence 
the gravitino mass since both $V_M^2$ and $V_M^3$ are zero.}
\be
\Lag_m^{\rm bulk} = -\frac{1}{4} \bar \Psi_a^\prime \gamma^{ab\dot 5}\tau_1\Psi_b^\prime 
\Big(Y^{\prime 2} \partial_5 Y^{\prime 3} - Y^{\prime 3} \partial_5 Y^{\prime 2}\Big)
= -\frac 12 \partial_5\alpha_1 \bar \Psi_a^\prime \gamma^{ab\dot 5}\tau_1\Psi_b^\prime \;.
\label{mass1}
\ee
The value of $Y^{\prime i}$ at the fixed-points is more subtle. At first sight 
$Y^{\prime 2}=\cos 2\alpha_1$ would seem continuous, even though $\alpha_1$ flips sign
at the boundaries. Then one would conclude that, in eq.~(\ref{superboundary}), one should 
use $Y^{\prime 2}(0)=\cos\pi\eta_0$ and $ Y^{\prime 2}(\pi)=\cos\pi\eta_\pi$.
However this simple reasoning is incorrect. The point is that $Y^2$, strictly at the fixed 
points, is invariant under $\hbox{U}(1)_{T_1}$. This is because the orbifold projection breaks 
${\rm SU}(2)_R$ down to $\hbox{U}(1)_{T_3}$ and $\hbox{U}(1)_{T_1}$ is not active at the boundaries. 
Therefore, based on gauge invariance, we must impose $Y^{\prime 2}(0)=Y^{\prime 2}(\pi)\equiv 1$. 
This is equivalent to taking $\alpha_1(0)=\alpha_1(\pi) \equiv 0$, which makes qualitative 
sense since $\alpha_1$ is on average zero at the fixed points. $Y^{\prime 2}$ is discontinuous 
at the fixed points, even though, being even, it has the same limit when approaching the 
fixed points  from opposite sides. In the end we must  take $Y^{\prime 2}=1$ and $Y^{\prime 3}=0$ 
at the fixed points, even though in the bulk they rotate. The mass term induced by the 
boundary superpotential is then given by
\be
\Lag_m^{\rm bound} = \frac 12 P_\eta(x^5) \bar \Psi_a^\prime \gamma^{ab}\tau_2\Psi_b^\prime \;.
\label{mass2}
\ee 
The total mass term is found by adding (\ref{mass1}) and (\ref{mass2}). Using the fact that 
$i \tau_3 \gamma^{\dot 5}=1$ on the  gravitino at the boundary, one can verify that the 
contributions in (\ref{mass1}) that are localized at the boundaries exactly cancel (\ref{mass2}), 
and only a constant bulk mass term is left:
\be
\Lag_m = \frac{\eta}{2}\bar \Psi_a^\prime \gamma^{ab\dot 5}\tau_1\Psi_b^\prime \;.
\label{constmass}
\ee
The most appropriate basis of KK wave-functions is in this case obtained from the standard 
one through a $\hbox{U}(1)_{T_1}$ rotation that diagonalizes the constant bulk mass terms 
(\ref{constmass}):
\be
\xi_n^\prime(x^5) = \left (\matrix{e^{i\pi/4}\cos nx^5\cr -e^{-i\pi/4}\sin nx^5\cr}\right) \;.
\label{eigenmodes1}
\ee
This leads to the following mass eigenvalues:
\be
m_n=\frac{n+\eta}{R} \;.
\ee
Again, the mass of the lightest mode agrees with eqs.~(\ref{V1smooth}) and (\ref{noscalemass}). 
In the case where both auxiliaries $V_5^1$ and $V_5^2$ are turned on, the parameter describing 
the twisted gravitino spectrum becomes $|\eta+i\epsilon|\propto |V_5^1+iV_5^2|$.

If we rotate the eigenmodes back to the original gauge, their wave-functions become 
$\xi_n(x^5) = e^{i \alpha_1(x^5) \tau_1} \xi^\prime_n(x^5)$.
From this expression, and from the rule $\alpha_1(0)=\alpha_1(\pi)\equiv 0$, we deduce that 
$\chi^1$ not only has a cusp but it is truly discontinuous at the fixed points. This suggests 
that a derivation of the spectrum based on the operator eq. (\ref{massoperator}) assuming 
continuity of $\chi^1$ is flawed. Indeed under these assumptions we would get different 
eigenvalues that do not satisfy the periodicity under $\eta \to \eta+1$:
\be
m_n= \frac {n + \arctan \eta_0+\arctan \eta_\pi}R \;.
\ee
Notice also that in the singular basis it is the continuous combination of $\chi^1$ and 
$\chi^2$ that couples to the boundary. In ref.~\cite{Gherghetta} this point was missed: only 
$\chi^1$ was coupled in the computations. As a consequence, the coupling of the gravitini to 
matter at $x^5=0,\pi$ was weighted by the wave function factor $\cos \eta_{0,\pi} \pi$. 
This way if one were to repeat the computation of section 4 one would find that the supersymmetric 
cancellation is spoiled and that the scalar masses are UV divergent. Notice that this disaster 
would also survive the case $\eta_0=-\eta_\pi\not =0$ in which half of the supercharges are 
preserved (there is a killing spinor~\cite{Zucker}) and not even a finite scalar mass is tolerated.
We believe that our approach makes it clear how to avoid these errors.

\setcounter{equation}{0}
\section{One-loop effective action}
\label{calcolo}

We can now compute the one-loop correction to the K\"ahler potential.
We consider a `visible sector' consisting of a chiral multiplet $\Phi_0$ with
kinetic function $\Omega_0(\Phi_0,\Phi_0^\dagger)$ localized at $x^5 = 0$ and 
a `hidden sector' consisting of a chiral multiplet $\Phi_\pi$ with 
kinetic function $\Omega_\pi(\Phi_\pi,\Phi_\pi^\dagger)$ at $x^5 = \pi$. 
In the notation of section~\ref{sugra}, the 5D microscopic theory is 
described at tree level by the generalized kinetic function
\be
\Omega(x^5) = - \frac 3{2} + \Omega_0(\Phi_0,\Phi_0^\dagger) e_{\dot 5}^5\, \delta(x^5) 
+ \Omega_\pi(\Phi_\pi,\Phi_\pi^\dagger) e_{\dot 5}^5\, \delta(x^5 - \pi ) \;.
\ee
The 4D low-energy effective supergravity theory obtained by integrating out all the massive KK 
modes is then specified at leading order by $\Omega = \int_{-\pi}^\pi dx^5e_5^{\dot 5}\,\Omega(x^5)$, 
which leads to eq.~(\ref{Omega0}). As explained in section 2, one-loop diagrams involving the 
massive supergravity KK modes will induce a correction $\Delta \hat \Omega$ mixing the two sectors.
$\Delta \hat \Omega$ can be fully reconstructed by calculating the scalar potential 
$\Delta V$ in a background with $F_T\not =0$. For definiteness we can consider the case discussed 
in \ref{generateV2}, where $F_T=\pi V_5^2=2\pi\epsilon$ and the gravitino KK modes are already 
well behaved in the original basis.

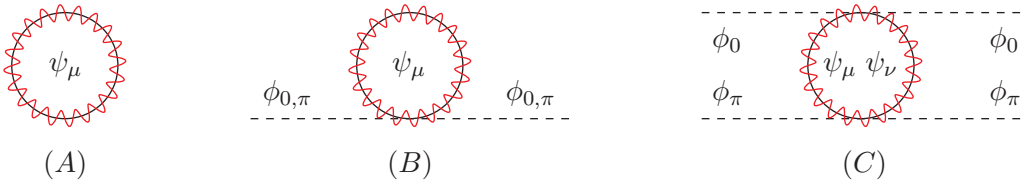
\begin{figure}[h]
\begin{center} 
\begin{picture}(440,80)(-20,-25)
\put(14,18){$\psi_{\mu}$}
\CArc(20,20)(20,0,360)
\SetColor{Red}
\PhotonArc(20,20)(20,0,360){3}{20}
\SetColor{Black}
\put(12,-20){$(A)$}
\put(95,7){$\phi_{0,\pi}$}
\put(187,7){$\phi_{0,\pi}$}
\put(144,18){$\psi_{\mu}$}
\DashLine(90,0)(210,0){3}
\CArc(150,20)(20,0,360)
\SetColor{Red}
\PhotonArc(150,20)(20,0,360){3}{20}
\SetColor{Black}
\put(142,-20){$(B)$}
\put(265,7){$\phi_\pi$}
\put(370,7){$\phi_\pi$}
\put(265,27){$\phi_0$}
\put(370,27){$\phi_0$}
\put(307,18){$\psi_{\mu}$}
\put(323,18){$\psi_{\nu}$}
\DashLine(260,0)(380,0){3}
\DashLine(260,40)(380,40){3}
\CArc(320,20)(20,0,360)
\SetColor{Red}
\PhotonArc(320,20)(20,0,360){3}{20}
\SetColor{Black}
\put(314,-20){$(C)$}
\end{picture}
\caption{\em Diagrams controlling $(A)$  Casimir energy,
$(B)$ radion-mediation and
$(C)$ brane to brane mediation of supersymmetry breaking.
\label{fig:ABC}}
\end{center}
\end{figure}

In the presence of a non-zero $F_T = 2 \pi \epsilon$, the component expansion for the $D$-term 
of the correction 
\be
\Delta \hat \Omega = \frac{A}{(T+T^\dagger)^2}+B
\frac {\Phi_0 \Phi_0^\dagger + \Phi_\pi \Phi_\pi^\dagger}{(T + T^\dagger)^3}
+ C \frac {\Phi_0 \Phi_0^\dagger \Phi_\pi \Phi_\pi^\dagger}{(T + T^\dagger)^4}
+ \cdots 
\label{DeltaOmegaABC}
\ee
leads to the scalar potential 
\be
\Delta V = - |F_T|^2\,\frac {\partial^2 \Delta \hat \Omega}{\partial(T + T^\dagger)^2}
= -\frac {3 A \epsilon^2}{2 \pi^2 R^4}
- \frac {3 B \epsilon^2}{2 \pi^3 R^5} \Big(|\phi_0|^2 + |\phi_\pi|^2\Big)
- \frac {5 C \epsilon^2}{4 \pi^4 R^6} |\phi_0|^2 |\phi_\pi|^2 + \cdots \;.
\label{O}
\ee
Each of the interactions in $\Delta V$ is contributed to by many diagrams, adding up to zero 
in the supersymmetric limit $\epsilon \rightarrow 0$. However, only the gravitino 
spectrum is affected by $\epsilon$; the graviton and graviphoton are unaffected by supersymmetry 
breaking. Therefore it is enough to compute the contribution of gravitino diagrams $\Gamma_\Psi(\epsilon)$. 
The potential is then simply given by $\Delta V(\epsilon) = \Delta V_\Psi(\epsilon) - \Delta V_\Psi(0)$. 
As discussed in section 4, all diagrams involving cubic gravitino vertices vanish at zero 
momentum, so that there is a single relevant diagram for each operator in $\Delta \Omega$, involving 
the quartic vertex, as depicted in Fig.~\ref{fig:ABC}.

Since supersymmetry is broken, all the gravitino modes are massive. For this reason it is convenient 
to work with Majorana spinors $\psi_i^\mu = (\chi_i^\mu,\bar \chi_i^\mu)^T$. The relevant interaction,
which is nothing but the standard field-dependent (localized) gravitino kinetic term, is then 
written as
\be
\Lag^{\rm int} = \frac 13 \Big[\Omega_0(\phi_0,\phi_0^*)\,\delta(x^5) 
+ \Omega_\pi(\phi_\pi, \phi_\pi^*)\, \delta(x^5 - \pi) \Big] \,
i e_4 \bar \psi_\mu \gamma^{\mu\nu\rho} \partial_\nu \psi_\rho 
\ee
In this case it is convenient to work in the unitary gauge $\psi_i^5 = 0$ to decouple completely the 
Goldstinos. The gravitino propagator for a mode of mass $m$ is then given by the ordinary propagator 
for a massive gravitino~\cite{VanNieuwenhuizen}:
\be
\langle \psi_\mu \bar \psi_\nu\rangle_n = 
\Big[\eta_{\mu \nu} - \frac {p_\mu p_\nu}{m_n^2} 
- \frac 13 \Big(\gamma_\mu - \frac {p_\mu}{m_n}\Big)
\Big(\gamma_\nu - \frac {p_\nu}{m_n}\Big) \Big]
\frac i{p\!\!\!/ + m_n} \label{proppsimaj} \;.
\ee
As in section~\ref{test}, the longitudinal part of the propagator is irrelevant for 
the amplitudes with vanishing external momentum that we are interested in. 

The diagrams that we have to compute consist of gravitinos propagating between interaction
vertices localized at the fixed-points. Therefore the  sums over virtual KK modes $n$ reconstruct
in position space the propagator between the 
two fixed-points. More precisely, each vertex occurring at the fixed-point located at 
$x^5 = k\pi$ comes with a wave-function factor $e^{i n k \pi}$, and the sum over KK modes 
in a propagator connecting two fixed-point separated by a distance $d$ will therefore be 
weighted by a factor $e^{i n d}$. After going to Euclidean space and performing the trace 
over spinor indices, all the diagrams can be reexpressed in terms of the basic quantity 
$G_d(p,\epsilon) = (2 \pi R)^{-1} \sum_n e^{i n d}/ (p - i\, m_n(\epsilon))$. 
The distance $d$ is $0$ when the propagation is from a fixed-point to itself, and $\pi$ 
when the propagation occurs instead from one fixed-point to the other. The relevant 
quantities are therefore
\bea
G_0(p,\epsilon) \a=\a \frac 1{2 \pi R} \sum_{n=-\infty}^\infty
\frac 1{p - i\, m_n(\epsilon)} = \frac 12 \coth \pi(p R + i \epsilon) \;, 
\label{G0} \\
G_\pi(p,\epsilon) \a=\a \frac 1{2 \pi R} \sum_{n=-\infty}^\infty
\frac {(-1)^n}{p - i\, m_n(\epsilon)}
= \frac 12\,{\rm csch}\, \pi(p R + i \epsilon) \;.
\label{Gpi}
\eea

\subsection{Models without localized kinetic terms}

Let us first examine the simplest situation in which the boundary terms $L_0$ and $L_\pi$
are set to zero in (\ref{lowest0}) and (\ref{lowestpi}). In this case, one can use the 
the standard bulk gravitino propagator to compute the diagrams of Fig.~\ref{fig:ABC}.

Consider first the vacuum diagram $A$. Its standard expression as the trace of the 
logarithm of the kinetic operator can be rewritten 
in terms of (\ref{G0}) thanks to an integration by parts which isolates the divergent 
$\epsilon$-independent part:
\bea
\Delta V_{\Psi}^A(\epsilon) \a=\a - 8\,\frac {1}{2} \sum_{n=-\infty}^\infty 
\int \frac {d^4p}{(2 \pi)^4} \ln [p^2 + m_n^2(\epsilon)] \nn \\
\a=\a {\rm Div.} + 2 \pi R \int \frac {d^4p}{(2 \pi)^4}\,
p\, \mbox{Re} \Big[G_0(p,\epsilon)\Big] \;.
\label{vac}
\eea
The total vacuum amplitude is given by $\Delta V_A(\epsilon) = \Delta V_\Psi^A(\epsilon) 
- \Delta V_\Psi^A(0)$, so that one is left with a finite momentum integral which is easily 
evaluated:
\bea
\Delta V_A(\epsilon) 
\a=\a \frac {3}{16 \pi^6 R^4}\,\Big[\mbox{Re}\,{\rm Li}_5(e^{2 \pi i \epsilon}) - \zeta(5) \Big]
\simeq - \frac {3 \zeta(3) \epsilon^2} {8 \pi^4 R^4} \;.
\eea
In the last step, we have used $\mbox{Re}\,{\rm Li}_5(e^{2 \pi i \epsilon}) 
\simeq \zeta(5) - 2 \pi^2 \zeta(3) \epsilon^2 + {\cal O}(\epsilon^4)$ in the limit 
$\epsilon \rightarrow 0$.
Comparing with eq.~(\ref{O}) we get
\be
A = \frac {\zeta(3)}{4\pi^2} \;.
\ee

Consider next the two-point function $B$. The diagram is easily evaluated, and the result 
in Euclidean space is given by the following expression:
\bea
\Delta V_\Psi^B(\epsilon) \a=\a \frac {4}{3(2 \pi R)}
\sum_{n=-\infty}^\infty \int \frac {d^4p}{(2 \pi)^4}
\frac {p^2}{[p^2 + m_n^2(\epsilon)]} \nn \\
\a=\a \frac 43 \int \frac {d^4p}{(2 \pi)^4}\,
p\, \mbox{Re} \Big[G_0(p,\epsilon)\Big] \;.
\eea
The total contribution $\Delta V_B(\epsilon) = \Delta V_\Psi^B(\epsilon) - \Delta V_\Psi^B(0)$ 
is finite and given by:
\bea
\Delta V_B(\epsilon) 
\a=\a \frac {1}{8 \pi^7 R^5}\,\Big[\mbox{Re}\, {\rm Li}_5(e^{2 \pi i \epsilon}) - \zeta(5) \Big] 
\simeq - \frac {\zeta(3) \epsilon^2} {4 \pi^5 R^5} \;.
\eea
Comparing with eq.~(\ref{O}), one extracts:
\be
B = \frac {\zeta(3)}{6\pi^2} \;,
\ee
in agreement with~\cite{Gherghetta}. \footnote{In ref. \cite{Gherghetta} the result depends on whether the 
constant superpotential is on the visible or hidden brane. This is because of their incorrect treatment 
of the gravitino wave function. We are comparing here with their formula for a superpotential at the hidden 
brane. In this case the gravitino field is smooth at the visible brane and the result of 
ref.~\cite{Gherghetta} correct.}

Consider finally the four-point function $C$. After some straightforward algebra, the
diagram can be simplified to: 
\bea
\Delta V_\Psi^C(\epsilon) \a=\a \frac {4}{9(2 \pi R)^2}
\sum_{n,n^\prime=-\infty}^\infty \int \frac {d^4p}{(2 \pi)^4}
\frac {p^2[p^2 - m_n(\epsilon) m_{n^\prime}(\epsilon)]}
{[p^2 + m_n^2(\epsilon)][p^2 + m_{n^\prime}^2(\epsilon)]}
(-1)^{n+n^\prime} \nn \\
\a=\a \frac 49 \, \int \frac {d^4p}{(2 \pi)^4}\,
p^2\, \mbox{Re} \Big[G_\pi(p,\epsilon)^2\Big] \;.
\eea
In this case, the momentum integral is finite even before subtracting the untwisted diagram 
$\Delta V_\Psi^C(0)$, because the loop involves propagation between separated fixed-points and therefore 
cannot shrink to a point. The final result is
\bea
\Delta V_C(\epsilon) 
\a=\a \frac {5}{48 \pi^8 R^6}\,\Big[\mbox{Re}\, {\rm Li}_5(e^{2 \pi i \epsilon}) - \zeta(5) \Big] 
\simeq - \frac {5 \zeta(3) \epsilon^2} {24 \pi^6 R^6} \;.
\eea
This yields:
\be
C = \frac {\zeta(3)}{6\pi^2} \;.
\ee

Since  $A,B,C$ are positive, if the operators in eq.~(\ref{DeltaOmegaABC}) are dominant
the radion potential is unbounded from below at $R\to 0$
and visible-sector scalars get negative squared masses for any $R$.

\subsection{Models with localized kinetic terms}

In the more general situation in which non-vanishing boundary terms $L_0$ and $L_\pi$ 
arise in (\ref{lowest0}) and (\ref{lowestpi}), the computation is more involved. 
In this case, one has to dress the diagrams of Fig.~\ref{fig:ABC} with insertions of the boundary 
kinetic terms for the bulk fields, and resum all of these. This is equivalent to compute 
the exact 1-loop effective potential as a function of $\Omega_0$ and $\Omega_\pi$. 
In order to perform this computation, it is crucial to use the projection operators defined in 
section~\ref{test} in order to simplify the tensor structure of the interaction and the propagator. 
Defining for convenience $\rho_{0,\pi} = \frac 13 \Omega_{0,\pi}(\phi_{0,\pi},\phi_{0,\pi}^*)$,
the scalar-scalar-gravitino-gravitino coupling can be written as:
\bea
\Lag^{\rm int} = \Big(\rho_0\, \delta(x^5) + \rho_\pi\, \delta(x^5-\pi)\Big)
\bar \psi_\mu\, p\!\!\!/\,(P_{3/2}^{\mu \nu} - 2 P_{1/2}^{\mu\nu}) \psi_\nu \;.
\eea
Since the longitudinal part of the gravitino propagator is irrelevant, it can be conveniently
chosen in such a way to reconstruct the projection operator $P_{3/2}$ in the polarization 
factor. By doing so, the gravitino propagators between two fixed-point separated by a distance 
$d=0,\pi$ can be written as 
\bea
\Delta_{d(\psi)}^{\mu\nu} \a=\a p\!\!\!/\, P_{3/2}^{\mu\nu}\, \Delta_d \;,
\eea
with:
\bea
\Delta_0 \a=\a \frac 1{2 \pi R} \sum_{n=-\infty}^\infty  
\frac {i}{p\!\!\!/\,(p\!\!\!/\, + m_n)} 
\label{Delta1} \;, \\
\Delta_\pi \a=\a \frac 1{2 \pi R} \sum_{n=-\infty}^\infty 
\frac {i(-1)^n}{p\!\!\!/\,(p\!\!\!/\, + m_n)} \;.
\label{Delta2} 
\eea
The Euclidean versions of these quantities reduce for $\epsilon=0$ to $-i/p$ times 
(\ref{G0}) and (\ref{Gpi}).

The effective potential is obtained by summing up all the independent diagrams with an arbitrary 
number of each type of insertion. This task is complicated by the fact the type of propagators 
to be used depends on the topology of the diagram, and not just on the number of each type 
of insertion. The easiest way to figure out the correct combinatoric is then to resum the two 
kinds of insertions successively. First one computes a dressed propagator that takes into account 
one type of insertion, say the insertion of $\rho_\pi$. Then one uses this propagator 
to compute diagrams with a given number of the other insertion (insertion of $\rho_0$) and 
finally resums the latter.

The effective gravitino propagator between two $\rho_0$ vertices corrected by insertions of 
$\rho_\pi$ vertices is easily computed by using a geometric resummation; the result can be 
written as $\Delta_{0(\psi)}^{\mu\nu}(\rho_\pi) = p\!\!\!/\, P_{3/2}^{\mu\nu}\, 
\Delta_0(\rho_\pi)$ with
\bea
\Delta_0(\rho_\pi) \a=\a 
\Delta_0 + \Delta_\pi \,(-i\rho_\pi)\,p^2\, \Delta_\pi 
+ \Delta_\pi \,(-i\rho_\pi)\,p^2\, \Delta_0 \,(-i\rho_\pi)\,p^2\, \Delta_\pi 
+ \cdots \nn \\
\a=\a  \Delta_0 - i\,\rho_\pi\,p^2\,\Delta_\pi^2\,
\Big(1 + i\,\rho_\pi\,p^2\,\Delta_0\Big)^{-1} \;. 
\eea
The full effective potential is then given by the sum of two pieces. The first is the 
sum of all the diagrams with at least one insertion of $\rho_0$, but computed with the 
dressed propagator $\Delta_0(\rho_\pi)$. The second is the effective potential at 
$\rho_0=0$, which corresponds to all the diagrams with only $\rho_\pi$ vertices and 
undressed propagator $\Delta_0$. The result is:
\bea
W_\Psi(\rho_{0,\pi}) \a=\a \frac 12 \int \! \frac {d^4p}{(2 \pi)^4} \sum_{k=1}^{\infty} 
\frac {(-i\,p^2)^k}k {\rm Tr} \Big[\Big(\,\rho_0\,P_{3/2}\, \Delta_0(\rho_\pi)\Big)^k 
+ \Big(\,\rho_\pi\,P_{3/2}\, \Delta_0\Big)^k\Big] \nn \\
\a=\a - P^{\,\mu}_{3/2\,\mu} \,\frac 12 \int \! \frac {d^4p}{(2 \pi)^4} {\rm Tr} \ln 
\Big[1 + i(\rho_0 + \rho_\pi)\,p^2\, \Delta_0 
- \rho_0\,\rho_\pi\, p^4\, \Big(\Delta_0^2 - \Delta_\pi^2\Big) \Big] \;.\hspace{20pt}
\label{eq:gravitino}
\eea
The trace over vector indices reduces therefore to $P^{\,\mu}_{3/2\,\mu} = 2$. 
The trace over spinor indices is less immediate, but can be easily performed as well. 
Introducing a matrix notation for the propagation between the two types of boundaries, 
and going to Euclidean space, the final result can be written in terms of the complex 
propagators (\ref{G0}) and (\ref{Gpi}) as
\bea
W_\Psi(\rho_{0,\pi}) 
\a=\a -{\rm Tr} P_{3/2}\,\frac 12 \int \! \frac {d^4p}{(2 \pi)^4} \, {\rm Re}\,\ln \det
\left(\matrix{1 - \rho_0\,p\, G_0 \a - \rho_0\,p\, G_\pi \smallskip \cr
-\rho_\pi\,p\, G_\pi \a 1 - \rho_\pi\,p\, G_0}
\right) \;,
\label{genpop}
\eea
where ${\rm Tr} P_{3/2} = 8$ is the total number of degrees of freedom and the determinant 
is now only as a $2 \times 2$ matrix. The structure of the result is therefore 
$V = \frac 12 {\rm Tr}\,{\rm ln}\,[1 - p\, M(\rho_{0,\pi})]$ where $M(\rho_{0,\pi})$ is a  
$2 \times 2$ matrix encoding the propagation between any pair of fixed-points weighted by the appropriate 
coupling $\rho_0$ or $\rho_\pi$. In more complicated situations with $N$ distinct fixed-points 
$i=1,\dots, N$ with couplings $\rho_i$, $M$ would generalize to the $N \times N$ matrix 
$M_{ij} = \rho_i G_{ij}$.

As in the simpler case analyzed in section 4, the above result is indeed the expected induced 
effective potential for a theory with twisted boundary conditions and localized kinetic terms. 
As already explained, the twisting influences only the mass of the gravitino modes, but not the 
strength of their couplings to the boundaries. For this reason, the main features of the computation 
are already captured by the untwisted case. The contribution of a single untwisted scalar degree of 
freedom with Lagrangian 
\be
\Lag_\varphi = \partial_M \varphi\, \partial^M \varphi 
+ \Big(\rho_0\, \delta(x^5) + \rho_\pi\, \delta(x^5-\pi) \Big)
\partial_\mu \varphi \, \partial^\mu \varphi
\ee
has been studied in~\cite{Ponton}. The result can be reobtained in an alternative and very simple way 
as a determinant. Using $f_n = i/(p^2-m_n^2)$, in terms of which the propagator (\ref{Delta1}) 
and (\ref{Delta2}) for $\epsilon = 0$ read simply $\Delta_0 = (2 \pi R)^{-1}\sum_n f_n$ and 
$\Delta_\pi = (2 \pi R)^{-1}\sum_n (-1)^n f_n$, one computes: 
\bea
W_{\varphi} (\rho_{0,\pi}) 
\a=\a \frac 12 \, {\rm ln}\, {\rm det}\, 
\Big[1 - \Big(\rho_0\,\delta(x^5) + \rho_\pi\,\delta(x^5-\pi)\Big)\, 
\frac {\partial_\mu \partial^\mu}{\partial_M \partial^M}\Big] \nn \\
\a=\a \frac 12 \int \! \frac{d^4p}{(2\pi)^4} {\rm ln}\,{\rm det}_{\rm KK} 
\Big[\delta_{n,n^\prime} + \frac {i\rho_0\, p^2}{2 \pi R} f_n
+ \frac {i\rho_\pi\, p^2}{2 \pi R} (-1)^{n+n^\prime} f_n\Big] \nn \\
\a=\a \frac 12 \int \! \frac{d^4p}{(2\pi)^4} {\rm ln}\, 
\Big[1 + \frac {i (\rho_0+\rho_\pi)\, p^2}{2 \pi R}\,\raisebox{2pt}{$\displaystyle{\sum_n}$}\,f_n
- \frac {4 \rho_0\,\rho_\pi\, p^4}{(2 \pi R)^2}\,\raisebox{2pt}{$\displaystyle{\sum_{n,n^\prime}}$}
f_{2n} f_{2n^\prime+1} \Big] \nn \\
\a=\a \frac 12 \int \! \frac{d^4p}{(2\pi)^4} {\rm ln}\, 
\Big[1 + i (\rho_0+\rho_\pi)\, p^2 \Delta_0 
- \rho_0\,\rho_\pi \, p^4 \Big(\Delta_0^2 - \Delta_\pi^2 \Big) \Big] \;.
\label{eq:spin0}
\eea
The infinite-dimensional KK determinant in the third step can be computed as before by considering 
finite-dimensional truncations. Comparing eq.~(\ref{eq:gravitino}) with eq.~(\ref{eq:spin0}), we see 
that the gravitino contributes indeed as $-8$ times a scalar. The result (\ref{genpop}) generalizes the 
results of~\cite{Ponton} to arbitrary boundary conditions. A similar computation for gauge fields 
can be found in~\cite{sss}.

The explicit expressions of the gravitino contribution to the vacuum energy (eq.~(\ref{vac})) and 
to the effective potential (eq.~(\ref{genpop})), as functions of the supersymmetry breaking parameter 
$\epsilon$, are given by
\bea
E_\Psi(\epsilon) \a=\a {\rm Div.} - \frac {1}{2 \pi^6 R^4}\, {\rm Re} \int_0^\infty \!\! dx\,x^3\,\ln 
\Big[\sinh (x + i \pi \epsilon)\Big] \;, \\
W_\Psi(\epsilon) \a=\a - \frac 1{2 \pi^6 R^4}\, {\rm Re} \int_0^\infty \!\! dx\,x^3\,\ln 
\Big[1 - (\alpha_0 + \alpha_\pi) x \coth (x + i \pi \epsilon) + \alpha_0\,\alpha_\pi x^2 \Big] \;,
\eea
in terms of the dimensionless parameters
\be
\alpha_{0,\pi} = \frac {\rho_{0,\pi}}{2 \pi R} = \frac {\Omega_{0,\pi}}{6 \pi R} \;.
\ee
The full effective action $\Delta V(\epsilon)$ is then obtained by subtracting the untwisted 
contribution, $\Delta V(\epsilon) = [E_\Psi(\epsilon) + W_\Psi(\epsilon)]- [E_\Psi(0) + W_\Psi(0)]$, 
and reads:
\bea
\Delta V(\epsilon) \a=\a - \frac 1{2 \pi^6 R^4}\,{\rm Re} \int_0^\infty \!\! dx\, x^3\,\ln 
\Big[1 - \frac {1 + \alpha_0 x}{1 - \alpha_0 x} \frac {1 + \alpha_\pi x}{1 - \alpha_\pi x}
e^{-2(x + i \pi \delta)} \Big]^{\delta=\epsilon}_{\delta=0} \;. 
\label{Gammat}
\eea
An alternative expression, which is particularly interesting in the case $L_{0,\pi} = 0$,
can be obtained by first expanding the logarithm in power series and then Taylor 
expanding the fractions around $\alpha_0 = \alpha_\pi = 0$ in a weak-field approximation. 
The first step is quite safe, but the second leads to an asymptotic series for the 
integrated result. Rescaling the integration variable, one finds in this way:
\bea
\Delta V(\epsilon) = \frac 1{2 \pi^6 R^4} \sum_{k=1}^\infty \frac 1{k^5} 
\Big(\cos 2 \pi k \epsilon - 1 \Big) \int_0^\infty \!\! dx\,x^3\,e^{-2x}\,
\Big[\sum_{p,q} \Big(\frac {\alpha_0 x}k\Big)^{|p|} 
\Big(\frac {\alpha_\pi x}k \Big)^{|q|}\Big]^k \;.
\label{Gammatot}
\eea
Using this power expansion, one finds ${\rm Li}_r$ functions of growing order $r$ for 
higher and higher order terms, which when expanded for $\epsilon \rightarrow 0$ yield 
$\zeta(r-2)$ functions. It is clear that for $L_{0,\pi} = 0$ all the infinite terms 
have the same sign, and working out the first few orders, one can easily check that 
$\Delta V(\epsilon) = \Delta V_A(\epsilon) + \Delta V_B(\epsilon) + \Delta V_C(\epsilon) 
+ \cdots$, reproducing therefore the diagrammatic computation.

Actually, it is possible to derive a closed integral form of the full one-loop 
correction (\ref{DeltaOmegaABC}) which encodes all the higher-order corrections as well. 
To do so, we rescale the integration variable by $1/(2 \pi R)$ and switch to the 
the $R$-independent quantities $\rho_{0,\pi}=\frac 13 \Omega_{0,\pi}$, to push the whole $R$-dependence 
of (\ref{Gammat}) into the exponential. This allows to relate in a simple way derivatives 
with respect to $\epsilon$ and derivatives with respect to $R$. At leading order in $\epsilon$, 
one finds; 
\bea
\Delta V(\epsilon) \a\simeq\a \frac {\epsilon^2}{\pi^2} \frac {\partial^2}{\partial R^2} 
\int_0^\infty \!\! dx\,x\,{\rm ln}\,
\Big[1 - \frac {1+\rho_0 x}{1-\rho_0 x} \frac {1+\rho_\pi x}{1-\rho_\pi x} e^{-4 \pi R x}\Big] \;.
\eea
Comparing this expression to eq.~(\ref{O}) with $F_T = 2 \pi \epsilon$ and restoring $M_5$,
one deduces finally the result anticipated in eq.~(\ref{DeltaOmega}) for the one-loop correction 
to the K\"ahler potential:
\bea
\Delta \hat \Omega \a=\a -\frac {9}{\pi^2} M_5^2 \int_0^\infty \!\!\!\! dx\, x\, \hspace{1pt}{\rm ln}
\Bigg[1 - \frac {1 + x\, \Omega_0 M_5^{-2}}{1 - x\, \Omega_0 M_5^{-2}}\,
\frac {1 + x\, \Omega_\pi M_5^{-2}}{1 - x\, \Omega_\pi M_5^{-2}}\,
e^{-6x(T + T^\dagger)M_5}\Bigg] \;.
\label{final}
\eea

\setcounter{equation}{0}
\section{Discussion}
\label{stab}

In this section we will study the visible soft terms arising from $\Delta \hat \Omega$.
The result will depend crucially on the mechanism by which the radion is stabilized.

\subsection{General analysis}

To start, we consider $R$ as a free parameter and deduce from $\Delta\hat\Omega$, eq.~(\ref{final}), an
explicit expression for the universal SUSY-breaking squared mass\footnote{We assume
that  the one loop correction $\Delta\hat\Omega$
negligibly renormalizes the tree level kinetic terms of matter fields, $\Omega_0$.
In such a situation $\Delta\hat\Omega$ induces a small universal trilinear term,
 $|A_0|\ll |m_0|$.}
$m_0^2$, which receives 
contributions from both $F_{\Phi_\pi}$ and $F_T$. Furthermore $m_0^2$ depends on 
the VEV of $\phi_\pi$, which might be non vanishing, and on $L_0$ and $L_\pi$ parametrizing 
the gravitational kinetic terms localized at the boundaries. We are particularly interested 
in studying in which cases $m_0^2$ is positive. Its explicit expression can be written as

\be
\label{mm0}
m_0^2 =\frac{\zeta(3)}{(4\pi)^2} \bigg[
- \frac{|F_{\Phi_\pi}|^2}{6 T^4 M_5^{6}} f_{\Phi_\pi\Phi_\pi}
- \frac{|F_{T}|^2}{T^5 M_5^{3}} f_{TT}
+ \frac{2 {\rm Re} [\phi_\pi F_T F_{\Phi_\pi}^*]}{3 T^5 M_5^6}
f_{\Phi_\pi T}\bigg] \;.
\ee
The functions $f_{\Phi_\pi\Phi_\pi}, f_{TT}, f_{\Phi_\pi T}$,
obtained by taking appropriate derivatives of $\Delta \hat \Omega$, depend on the 
dimensionless variables 
\be
\alpha_0 = \frac{\Omega_0}{6 M_5^3 T} = - \frac{L_0}{2 T} \;,\;\;
\alpha_\pi = \frac{\Omega_\pi}{6 M_5^3 T} = -\frac{L_\pi}{2 T} + 
\frac{|\phi_\pi|^2}{6 M_5^3 T} \;,\;\;
\beta = \frac{|\phi_\pi|^2}{6 M_5^3 T} \;.
\ee
The normalization has been chosen in such a way that 
$f_{\Phi_\pi\Phi_\pi} =  f_{TT} = f_{\Phi_\pi T}=1$ in the minimal 
case when all their three arguments vanish. One easily finds:
\bea
f_{\Phi_\pi\Phi_\pi} \a=\a 
\frac{2}{3\,\zeta(3)} \int_0^\infty \!\! dx \,x^3\,
\frac{(1+x^2\alpha_0(\alpha_\pi - 2\beta))\sinh x
- x(\alpha_0 + \alpha_\pi - 2\beta)\cosh x}
{[(1 + x^2 \alpha_0\alpha_\pi)\sinh x 
- x(\alpha_0 + \alpha_\pi)\cosh x]^3} \;, \hspace{10pt} \\
f_{TT} \a=\a 
\frac{1}{3\,\zeta(3)} \int_0^\infty \!\! dx \,x^4\,(1 - x^2\alpha_\pi^2)
\frac{(1+x^2\alpha_0\alpha_\pi)\cosh x
- x(\alpha_0+\alpha_\pi)\sinh x}
{[(1 + x^2 \alpha_0\alpha_\pi)\sinh x 
- x(\alpha_0 + \alpha_\pi)\cosh x]^3} \;, \hspace{10pt} \\
f_{\Phi_\pi T} \a=\a 
\frac{1}{3\,\zeta(3)} \int_0^\infty \!\! dx \,x^4\,
\frac{(1+x^2\alpha_0\alpha_\pi)\cosh x
- x(\alpha_0+\alpha_\pi)\sinh x}
{[(1 + x^2 \alpha_0\alpha_\pi)\sinh x 
- x(\alpha_0 + \alpha_\pi)\cosh x]^3} \;.
\eea
In the minimal case $L_0 = L_\pi = \phi_\pi = 0$, $m_0^2$ is negative.
In the presence of $\phi_\pi \neq 0$, but still keeping $L_{0,\pi} = 0$,
the third contribution to $m_0^2$ in eq.~(\ref{mm0}) can be positive, but it is
competitive with the first two only if $|\phi_\pi|^2\sim M_5^3T$. 
This situation is however  unphysical as it leads to an instability:
$\phi_\pi$ induces a negative localized kinetic term for the gravitational 
multiplet. For such large value of $\phi_\pi$ there is a ghostlike KK mode 
with a small tachyonic mass squared $m^2\sim -1/R^2$. To avoid manifest problems, 
we should take such a low UV cut-off $\sim 1/R$ for our 5D supergravity, that the
5D description itself is of no use. Therefore we do not consider this case.

In the presence of $L_{0,\pi} > 0$ such that
the localized kinetic terms are positive (i.e.\ $\alpha_{0,\pi}<0$)
$f_{\Phi_\pi\Phi_\pi}$ remains positive. Therefore, 
pure brane-to-brane mediation gives a negative contribution to $m_0^2$, 
corresponding to the term proportional to $|F_{\Phi_\pi}|^2$.
On the contrary $f_{TT}$
becomes negative for large enough $L_\pi$ (the precise value depends on $L_0$).
Therefore $m_0^2$ can be positive if the dimensionless quantity
\be
y=\frac{|F_{\Phi_\pi}|^2\,T}{|F_T|^2 M_5^3}
\ee
is small enough, i.e.\ in the presence of a radion-mediated contribution. 
Notice that generically we expect $|F_{\Phi_\pi}|^2/M_5^3T\sim |F_T|^2/T^2 \sim m_{3/2}^2$ 
so that $y\sim 1$ and radion mediation competes with brane-to-brane
mediation. In specific models things can however be different.

\begin{figure}[t]
$$\includegraphics{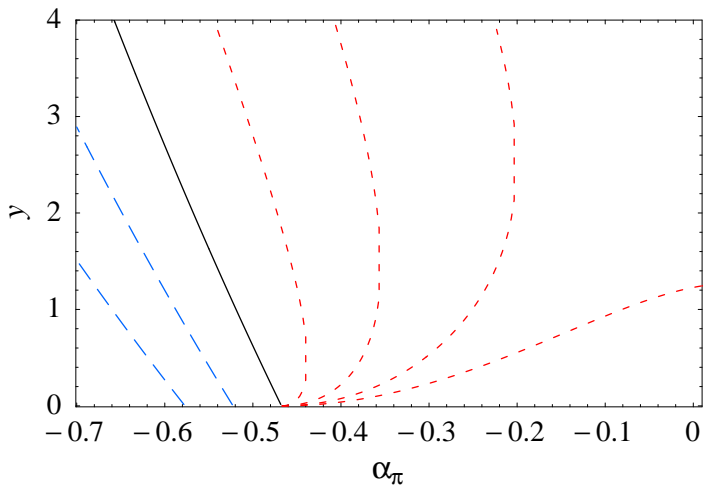}$$
\caption{\label{fig:mm0}\em $m_0^2$ can be positive at the left of the 
various lines, which correspond to representative values of $\alpha_0$
and $\beta$. The solid line corresponds to $\alpha_0=\beta = 0$,
whereas the blue long-dashed and red short-dashed lines describe 
situations with $\alpha_0 \neq 0$ and $\beta \neq 0$ respectively.}
\end{figure}

The situation is illustrated in Fig.~\ref{fig:mm0}: $m_0^2$ can be positive 
at the left of the various lines. The continuous line corresponds to 
$\alpha_0=\beta=0$. The blue long-dashed lines show how the boundary 
$m_0 = 0$ shifts when a non-zero $\alpha_0=\{-1/6,-1/2\}$ is turned on, while 
keeping $\beta=0$. Finally, the red short-dashed lines show how the boundary 
$m_0 = 0$ shifts when a non zero $\beta=\{1/12,1/6,1/4,1/3\}$ is turned on, 
while keeping $\alpha_0 = 0$. In the last case, as a consequence of the last 
term in eq.~(\ref{mm0}), $m_0^2$ can be positive even when the total localized 
kinetic terms vanish, $\alpha_{0,\pi} = 0$.

In conclusion $m_0^2$ is usually negative, but the radion-mediated 
contribution can make it positive in two basic circumstances:
1) if the gravitational multiplet has a sizable kinetic term localized 
on the hidden brane;
2) if the SUSY-breaking hidden sector field has a sizable VEV 
$\phi_\pi$.

A phenomenologically acceptable sparticle spectrum can be obtained
if some other effect generates supersymmetry breaking masses for gauginos.
We will later discuss the specific case of anomaly mediation,
which is the most natural candidate within the scenario we are considering.

In general, RGE effects induced by gaugino masses
can make scalar masses positive at low energy,
even starting from a negative $m_0^2$ at some high scale $\sim1/R$.
If scalar particles will be discovered,
extrapolating their masses up to high energies
one could try to identify a universal brane-to-brane contribution.
We remark that squared scalar masses can be negative at high energies:
this instability induces vacuum decay with a negligibly slow rate
(the thermal evolution of the universe can naturally select the metastable physical vacuum).

The low energy physical sfermion masses might contain non-SM sources of flavor and 
CP violation. In unified theories or in presence of large neutrino Yukawa couplings,
RGE corrections imprints detectable extra sources of flavor violations in scalar 
masses (see e.g. \cite{RGE}).
Beyond these effects, we expect that brane-to-brane mediation itself does not give an 
exactly flavor universal $m_0^2$ because gravity becomes flavor universal only at low 
energy, but in general violates flavor around the Planck scale.
In fact, the effective supergravity Lagrangian describing matter terms might contain
dimension 6 terms like e.g.\  kinetic terms with extra derivatives $(\partial/\Lambda_5)^2$ 
and flavor breaking coefficients. $\Lambda_5$ is the unknown energy at which
new quantum gravity phenomena not accounted by general relativity set in. 
Na\"{\i}ve dimensional analysis suggests $\Lambda_5\circa{<} 4\pi M_5$,
with approximate equality holding if quantum gravity is strongly coupled~\cite{nda}. 
In absence of a predictive theory of quantum gravity and of flavor, 
we cannot go beyond these semi-quantitative expectations.

Since brane-to-brane mediation is dominated by loop energies $E\sim 1/\pi R$,
higher-dimensional operators are expected to give
small flavor-breaking corrections to the squared masses proportional 
to the factor $\delta \sim 1/(\Lambda_5 \pi R)^2$.
If brane-to-brane mediation is used to solve the problems of anomaly mediation,
the discussion below eq.~(\ref{AM}) suggests $\delta\circa{>} 1/(4\pi)^{10/3}$.
On the experimental side, $\mu\to e\gamma$ and $\epsilon_K$
give the strongest bounds, $\delta \circa{<}10^{-3}$ for sfermion masses of a few 
hundreds of GeV~\cite{Masiero}.

\medskip

We will now discuss two different scenarios of radion stabilization.
We will focus on the case $\phi_\pi=0$ (or better $\phi_\pi \ll M_5$), 
suggested by a strongly coupled hidden sector. In this case the 4D Planck 
mass is given by
\be
M_P^2=M_5^3\left({\rm Re}T+L_0+L_\pi\right) \;.
\label{4dplanck}
\ee

\subsection{Luty--Sundrum model}

In ref.~\cite{Luty} the superpotential of the effective low energy theory was
given by
\be
P_{\rm eff}=\Lambda^2 \Phi_\pi +\frac{1}{16\pi^2}
\left(\Lambda_1^3+\Lambda_2^3e^{-a \Lambda_2 T/\pi}\right) \;.
\label{gaugecond}
\ee
The first term is just the standard O'Raifertaigh superpotential of the hidden sector. 
The second and third terms are generated by gaugino condensation of respectively a gauge 
group on the boundary and in the bulk\footnote{$\Lambda_{1,2}$ are the strong 
interaction scales according to NDA, $a$ is of order 1.}. Their r\^ole is to stabilize the 
radion, and also to allow to fine tune the 4D cosmological constant to zero. A discussion of 
the minimization of the potential is given in ref.~\cite{Luty}. One crucial remark that 
simplifies the discussion is that the $\Phi_\pi$ sector breaks supersymmetry already 
in the flat limit $M_5\to\infty$. Assuming that the superpotential for $\Phi_\pi$ originates from 
some strong 4D dynamics at the scale $\Lambda$ we have that the $\Phi_\pi$-dependent part in 
$\Omega_\pi$ has the form $\Phi_\pi\Phi_\pi^\dagger F(\Phi_\pi\Phi_\pi^\dagger/\Lambda^2)$ 
(see for instance ref.~\cite{Izawa:1997gs}). Then the scalar $\phi_\pi$ has a fairly large 
mass $\sim \Lambda\gg m_{3/2}\sim \Lambda^2/M_P$, so it can be integrated out before studying 
the radius potential. The only light mode in the $\Phi_\pi$ multiplet is the fermion, which 
contains a component of the eaten Goldstino. For the purpose of our discussions it is useful 
to briefly recall the resulting relations among the various parameters and VEVs. Cancellation 
of the 4D cosmological constant requires to tune 
\be
\frac{\Lambda_1^3}{(4\pi)^2}\sim \Lambda^2 M_P 
\ee
where the effective Planck mass $M_P$ is given by eq.~(\ref{4dplanck}), while the relevant VEVs are
\bea
F_{\Phi_\pi} \a\sim\a \Lambda^2 \;,\quad\quad \quad 
F_{S_0} \sim \frac{\Lambda_1^3}{(4 \pi M_P)^2}\sim \frac{\Lambda^2}{M_P}\sim m_{3/2} \;, \\
\frac{F_T}{T} \a\sim\a \frac{\pi F_{S_0}}{\Lambda_2\, T} \;,\quad\quad\quad 
\frac{\Lambda_2\, T}{\pi}\sim 3\ln\frac{\Lambda_2}{\Lambda_1} \;.
\eea
From the equation for $T$ it follows that its natural value is small. Indeed $\Lambda_2$ represents 
the strongly interacting scale of a bulk gauge theory, so it is natural to expect $\Lambda_2$ not 
much below the quantum gravity scale $\Lambda_5 \sim M_5\pi $. On the other hand, perturbative 
control of the 5D theory requires $\Lambda_2 T/\pi $ somewhat bigger than 1. In order to have 
scalar masses that are positive and comparable to gaugino masses, two conditions must be necessarily 
satisfied. One is that the anomaly mediated mass be comparable to the radion mediated mass
(second term in eq.~(\ref{mm0})). Using the above equations this condition reduces to
\be
\left (\frac{\Lambda_2 T}{\pi}\right )^2 (M_5 T)^3=(\Lambda_2 R)^2(\Lambda_5 R)^3
\sim \frac{\zeta(3)}{\pi^2}\left (\frac{4\pi}{g}\right )^4.
\ee
There is a window for which both gravity and the bulk gauge theory are (to a good extent) perturbative 
at the compactification radius. The other condition is that $m_0^2$ itself, eq.~(\ref{mm0}), be positive.
From the above minimum conditions we have
\be
y \sim \frac{\Lambda_2^2 T^2}{\pi^2} \left(1 + \frac {L_0}{T} + \frac {L_\pi}{T}\right).
\ee
This equation describes a line (not shown) in the plane of Fig.~\ref{fig:mm0}. 
For small enough $\Lambda_2 T/\pi$,
but parametrically still bigger than 1, the line will cross the black curve allowing positive masses. 
Considering the best case $L_0\sim 0$, we find that the crossing point is roughly at 
$L_\pi/T\sim \Lambda_2 T/\pi$: to safely remain in the perturbative domain of the bulk
gauge theory, the boundary contribution to $M_P^2$ should be hierarchically bigger than the 
ordinary 5D one, ($L_\pi\gg T$).

The fact that we can squeeze our parameters to get positive masses does not yet mean that we can build
a definite realistic model, in which all flavor violating contributions to the soft masses
are sufficiently small. As we have seen the parameters $1/\Lambda_2R$ and $1/\Lambda_5R$ controlling higher
order effects are not too small. Moreover for a large localized kinetic term (say $L_\pi$), gravity 
becomes strongly coupled at a fairly low scale scale $\sqrt {M_5/L_\pi}\ll M_5$~\cite{lpr}.
It would be interesting to make a thorough investigation.

\subsection{Radius stabilization through localized kinetic terms}

We will now consider the case in which the superpotential does not depend on $T$.
This is essentially the case we have considered in our calculation of the effective potential, 
and it is straightforward to include the effects of $F_{\Phi_\pi}$. We are then in the 
genuine no-scale scenario, in which $T$ is an exact flat direction at tree level. The 
effective potential we have calculated is a generalized Casimir energy lifting this 
flatness and one can ask if it can also stabilize the radius at some finite value. 
This issue has already been studied by Ponton and Poppitz~\cite{Ponton}, who have 
shown that appropriate localized kinetic terms for the bulk fields
lead to a modified Casimir energy with a stable minimum. This result is 
easy to understand. The boundary kinetic operator introduces a length scale $L$ in 
the theory. Therefore the Casimir energy $\sim 1/R^4$ is modified into $F(L/R)/R^4$, 
for which stationary points are possible at $R\sim L$. This effect of boundary terms 
is analogous to the one which we have already studied for the scalar masses.

In the presence of non-vanishing $F_T$ and/or $F_{\Phi_\pi}$ at tree level, the 
effective radion potential from (\ref{final}) is
\be
V(T) = -\frac {\partial^2 \Delta \hat \Omega}{\partial(T + T^\dagger)^2} \, |F_T|^2
- \frac {\partial \Delta \hat \Omega}{\partial{\Omega_\pi}}\, |F_{\Phi_\pi}|^2 \;.
\ee
Notice that at tree level we have $F_{S_0}=0$. This implies that anomaly mediated 
masses vanish at tree level in the gravitational interactions. $F_{S_0}=0$ is also 
associated, through the specific no-scale form of $\Omega_{\rm cl}$, to a vanishing 
contribution to the vacuum energy from the radion sector. At tree level the vacuum 
energy equals $|F_{\Phi_\pi}|^2>0$. The inclusion of the one loop correction 
$\Delta \Omega$ modifies this state of things.

The equation of motion of $F_T$ leads to 
\be
F_{S_0}=\frac{2}{3M_5^3}\frac {\partial^2 \Delta \hat \Omega}{\partial (T + T^\dagger)^2} 
\, F_T \sim \frac{\zeta(3)}{\pi^2 (M_5T)^3} \frac{F_T}{T} \;.
\label{Fcomp}
\ee
In the last equality we made a simple dimensional estimate, based on the assumption 
that the boundary kinetic terms introduce just one length scale (say $L_\pi$) which 
coincides with $T$ (see discussion below). Not surprisingly we find that $F_{S_0}$ 
is suppressed with respect to its natural scale just by the gravitational loop
expansion coefficient $\alpha_5=1/\pi^2(M_5T)^3$. The anomaly mediated gaugino 
masses are therefore similarly suppressed, as we will better discuss below.

\begin{figure}[t]
\begin{center}
\includegraphics[width=0.9\textwidth]{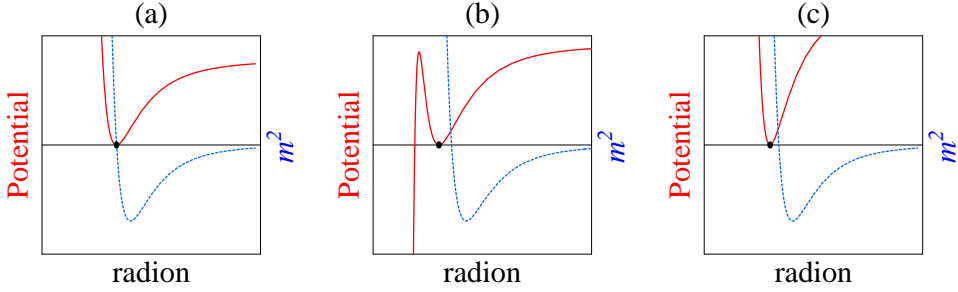}
\end{center}
\vskip -5pt
\caption{\em The radion potential $V(R)$ (red solid line) and the soft mass $m^2_0(R)$ 
(blue dashed line) in cases (a)  $L_0=0$, $L_\pi> 0$ (b) $0<L_0<L_\pi$ (c) $L_0=0$, $L_\pi> 0$ 
in presence of extra bulk fields.}
\label{figpot}
\end{figure}

At its minimum $V(T)$ is negative and the parameters can be tuned so that this 
contribution to the cosmological constant cancels the tree level contribution from 
$\Phi_\pi$. This cancellation leads to the following relation (tuning) between 
$F_T$ and $F_{\Phi_\pi}$
\be
F_{\Phi_\pi}\sim \frac {F_T}{\pi  T^2}\quad \rightarrow\quad 
y\sim \frac{1}{\pi^2 (M_5T)^3} \;.
\ee
Therefore in the perturbative regime we have $y\ll 1$ and radion mediation dominates 
the gravity induced scalar mass in eq.~(\ref{mm0}).

\medskip

We now consider various possibilities for the boundary kinetic terms, always 
assuming $\phi_\pi=0$.

\begin{itemize}
\item[(a)] $L_0=0$, $L_\pi>0$.

In this case, $V(T \rightarrow \infty) \rightarrow  0^-$ and $V(T \rightarrow 0) 
\rightarrow +\infty$ so that there is an absolute minimum at finite $T\sim L_\pi$. 
The radius can thus be made strictly stable. The soft mass $m^2_0(T)$ can be positive 
or negative, depending on $T$. But $m^2_0$ vanishes at the minimum of the potential, 
since it is given by $m^2_{0}(T) = -V^\prime(T)/6 \pi$. The situation is illustrated 
in Fig.~\ref{figpot}a. As already mentioned above, also the leading anomaly mediated 
masses vanish. In this case, to fully calculate the sparticle masses we should consider 
one extra gravitational loop for each quantity. We would then find that while the MSSM 
gauginos have mass $m_{1/2} \sim \alpha\alpha_5 |F_T|/T$, the scalars have a bigger 
mass $m^2_0\sim \alpha_5^2 (|F_T|/T)^2$, unless extra cancellations 
occur in the 2-loop gravitational contribution. Moreover there non-universal contribution to the scalar
masses, could be as important as the purely gravitational one.

\item[(b)]  $L_\pi\gg L_0>0$.

Now there is still a local minimum allowing for a metastable situation. The induced 
soft mass squared can become positive, and grows with $L_0$. Increasing $L_0$ lowers 
however the barrier hiding the true minimum, and for some critical value of 
$L_0 \sim  L_\pi$, the local minimum disappears and the potential becomes unstable. 
The situation is illustrated in Fig.~\ref{figpot}b. In this case 
$m_{1/2}\sim \alpha\alpha_5 |F_T|/T$, while $m_0^2\sim \alpha_5(|F_T|/T)^2$ arises 
already at 1-loop. So although  the tachyons can be avoided, the resulting model is 
phenomenologically quite unattractive, as a huge tuning ${\cal O}(\alpha^2\alpha_5)$ must 
be made in order to keep $M_Z^2\lsim m_{1/2}^2$ \cite{FT1} (multi-TeV universal scalars 
can be obtained by fine-tuning just the top Yukawa coupling \cite{FT2}).

\item[(c)] $L_0=0$, $L_\pi>0$, with extra bulk matter.

It is possible to add extra bulk fields that do not couple to ordinary matter and affect 
therefore only $V(R)$. Vector multiplets and hypermultipets give a contribution to $V(R)$ equal 
respectively to $\frac 12$ and $-\frac 12$ that of the supergravity multiplet. 
Introducing then $n_V$ vector multiplets and $n_H$ hypermultiplets with boundary kinetic 
terms that are independent from those of the supergravity multiplet, one can deform the 
effective potential to change the value of the radius at the minimum and therefore the 
value of the soft mass. If the localized kinetic terms of the new multiplets coincide with 
those of the supergravity multiplet, the minimum is not changed and $m_0^2$ remains 
zero. If they are smaller (bigger), then the minimum is shifted to lower (higher) $R$ for 
$n_V > n_H$ and higher (lower) $R$ for $n_V < n_H$, leading respectively to a positive 
(negative) and negative (positive) $m_0^2$ in the stable vacuum. An example with $L_0=0$, 
$L_\pi>0$ and extra bulk fields is illustrated in Fig.~\ref{figpot}c. Again, this scenario 
leads to a nice stable vacuum and positive $m_0^2$, but very light gauginos like in case b) 

\end{itemize}

\setcounter{equation}{0}
\section{Conclusions}
We studied gravity-mediated brane-to-brane supersymmetry breaking. We considered the simplest 
setup with one flat extra dimension, assumed to be a segment $S_1/\mathbb{Z}_2$ of length $\pi R$ with 
the `visible' MSSM fields localized at one boundary, and supersymmetry broken at the other 
`hidden' boundary. In this set-up, supersymmetry breaking is transmitted to MSSM fields by two 
different minimal effects: anomaly mediation (which gives gaugino masses and negative squared slepton 
masses), and one loop supergravity diagrams like the one depicted in Fig.~1 
(which gives an extra contribution to scalar masses).

We have computed this second contribution. Even if supergravity is plagued by UV divergences,
locality implies that Fig.~1 is finite and dominated by particles with energy $E\sim 1/R$.
Since gravity is flavor universal in the infrared, Fig.~1 induces a universal scalar soft 
mass squared $m_0^2$. Unknown UV effects give extra contributions suppressed by 
$\delta \sim 1/(M_5 R)^2$ which presumably break flavor. While the overall coefficient 
of $m_0^2$ depends on the 5D Planck mass $M_5$ and on $R$, its sign is strongly constrained.

Knowing that many contributions must cancel as demanded by 
supersymmetry we only needed to compute one Feynman graph,
plotted in fig.~\ref{fig:ABC}, which involves only one non trivial supergravity ingredient: 
the 4D coupling between two gravitinos and two scalars. 
In the first part of our paper we verified that all the rest works as expected.
Starting from the Lagrangian for off-shell 5D supergravity with localized 4D fields, we derived a 
much simpler partially on-shell formulation which can be conveniently used in loop computations.
(We also presented another less convenient formulation in which powers of $\delta(0)$ arise in intermediate steps).
We verified how supersymmetric cancellations really happen. Although not strictly 
necessary for our computation, we discussed these issues in great detail correcting in various 
ways previous attempts. We calculated the full 1-loop threshold correction
to the effective K\"ahler potential at the compactification scale. 
We have done this calculation indirectly.
We first computed the 1-loop effective potential in a convenient and consistent supersymmetry breaking
background, where constant superpotentials are placed at the boundaries. 
This set up corresponds to the well known Scherk-Schwarz mechanism. 
Secondly, we have reconstructed the full K\"ahler potential by solving a simple differential equation. 
The computation is fairly simple as it reduces to the class of gravitino loops. 
However in order to fully secure our result we had to tackle the puzzle
posed by the singular gravitino wave functions. These are a well know feature of models with boundary superpotentials
and can lead, if not properly treated, to ambiguities in loop computations. 
We have explained a simple procedure, based on
invariance under the local $\hbox{SU}(2)_R$ of the off-shell theory, to properly define the singular quantities and
obtain consistent results.

\smallskip

In the most minimal case we find $m_0^2 < 0$
(the same result has been obtained by Buchbinder, Gates, 
Goh, Linch III, Luty, Ng and Phillips using $N=1$ supergraph techniques~\cite{Lutyetal}). 
A positive $m_0^2$ arises in two basic 
circumstances: $i)$ if a substantial part of the 4D graviton kinetic energy comes from
terms localized on the hidden brane; $ii)$ if supersymmetry breaking fields localized 
on the hidden brane have a Planck-scale VEV. In both cases the radion superfield contributes 
to the mediation of supersymmetry breaking in an important way.

Finally, we studied how anomaly mediation and brane-to-brane effects may co-operate to give 
an acceptable sparticle spectrum --- a goal that neither of the two mechanisms reaches by 
itself. The two effects are comparable when the radius $R$ of the extra dimension is such 
that flavor-breaking higher order effects are suppressed by $\delta \circa{>} 1/(4\pi)^{10/3}$. 

We considered two possible concrete mechanisms of radius 
stabilization. In the first, where  the radion is stabilized by a superpotential generated by gaugino
condensation of a bulk gauge theory, we find that by stretching our parameters a little bit we can obtain 
positive scalar masses comparable to gaugino masses. This is a necessary requirement to construct a fully successful
model. To do so we must tackle the $\mu$-problem and carefully study all possible sources of flavor violation ---
a task that may be worth future work. The second scenario corresponds to a standard no-scale model in which
the radion is stabilized by the quantum corrections to the K\"ahler potential. Although this second
scenario presents peculiar phenomena, like the scalar masses vanishing exactly at the potential minimum in the
simplest realization, it has the phenomenological drawback of giving too small gaugino masses.

 We conclude listing a few related  questions,
 not addressed here because we do not know the precise answer.
What happens in a warped extra dimension?
In more than one extra dimension?
If matter is localized on fluctuating branes away from orbifold fixed points?

\section*{Acknowledgements}

We thank L.~Andrianopoli, I.~Antoniadis, S.~Ferrara, F.~Feruglio, F.~Girardello, G.~Giudice, C.~Kounnas, 
R.~Leigh, M.~Luty, A.~Riotto, M.~Serone, L.~Silvestrini, R.~Sundrum, A.~Zaffaroni and F.~Zwirner for 
useful discussions. This research was partly supported by the European Commission through a Marie Curie 
fellowship and the RTN research network HPRN-CT-2000-00148.

\end{document}